\documentclass[%
reprint,
superscriptaddress,
amsmath,amssymb,
aps,
pre,showpacs,
]{revtex4-1}
\usepackage{natbib}
\usepackage{subfigure}
\usepackage{graphicx}
\usepackage{dcolumn}
\usepackage{bm}
\usepackage{epstopdf}
\usepackage{epsfig}
\usepackage[colorlinks = true,
linkcolor = blue,
urlcolor  = blue,
citecolor = blue,
anchorcolor = blue]{hyperref}
\usepackage{url}
\usepackage[usenames,dvipsnames]{color}
\usepackage{xcolor}
\colorlet{RED}{red}

\usepackage{soul}

\DeclareMathOperator{\sech}{sech}

\begin{document}
	
	
	\title{Homoclinic snaking in the discrete Swift-Hohenberg equation}
	
	\author{R. Kusdiantara}%
	\email{rkusdi@essex.ac.uk}
	\affiliation{%
		Department of Mathematical Sciences, University of Essex, Wivenhoe Park, Colchester CO4 3SQ, United Kingdom
	}%
	\affiliation{Centre of Mathematical Modelling and Simulation, Institut Teknologi Bandung, 1st Floor, Labtek III, Jl.\ Ganesha No.\ 10, Bandung, 40132, Indonesia}
	
	\author{H. Susanto}
	\email{hsusanto@essex.ac.uk}
	\affiliation{%
		Department of Mathematical Sciences, University of Essex, Wivenhoe Park, Colchester CO4 3SQ, United Kingdom
	}%

	\date{\today}
	
	\begin{abstract}
		We consider the discrete Swift-Hohenberg equation with cubic and quintic nonlinearity, obtained from discretizing the spatial derivatives of the Swift-Hohenberg equation using {central} finite {differences}.
		We investigate the discretization effect on the bifurcation behavior, where we identify three regions of the {coupling parameter}, i.e.,\ strong, 
		weak, and intermediate coupling. 
		Within the regions, the discrete Swift-Hohenberg equation behaves either similarly or differently from the continuum limit.
		In the intermediate coupling region, multiple Maxwell points can occur for the periodic solutions and may cause irregular snaking and isolas. Numerical continuation is used to obtain and analyze localized and periodic solutions for each case.
		Theoretical analysis for the snaking and stability of the corresponding solutions is provided in the weak {coupling} region. 
	\end{abstract}
	
	\maketitle
	
	
	\section{Introduction}

	Homoclinic snaking in nonlinear dynamical systems is a snaking structure of the bifurcation curve {for spatially} localized states, which are homoclinic orbits in the phase space, in a parameter plane between a control parameter against, e.g., the norm of the states \cite{Woods2006}. 	A standard model for pattern formation and the commonly studied equation for homoclinic snaking is the Swift-Hohenberg equation with cubic and quintic nonlinearity \cite{Knobloch2008,Dawes2010,Purwins2010}, that models a physical problem of fluid having thermal fluctuation near Rayleigh-Bernard instability \cite{Swift1977,Getling1998}. 
	 The snaking structure has (possibly infinitely) many turning points, i.e.,\ saddle-node bifurcations, forming the boundaries of the snaking region \cite{Knobloch2008}. In spatially continuous systems, the localized structures can appear as a result of bistability between a uniform state and a periodic state around the uniform state itself. Generally, the two states are connected by a front which can drift in one direction. However, at a specific parameter value known as Maxwell point, the front has no preference between the two states which occurs when they have the same energy \cite{Budd2001,Hunt2000}. {Combining} two fronts back to {back forms} a localized state that can make a snaking structure in its bifurcation curve.	
	 The phenomenon has been studied theoretically in, e.g., \cite{Beck2009} that predicts the presence of snakes and ladders, \cite{Budd2005,Kozyreff2006} that analyze localized periodic patterns using multiple scale expansions, \cite{Burke2006,Burke2007,Burke2007a} that provide thorough numerical continuations of homoclinic snaking in the Swift-Hohenberg equation, and \cite{Clerc2005} that studies localized patterns as particle-type solutions (see also \cite{Dawes2010} for a short review of coherent structure emergence based on localized {structures}). Homoclinic snaking has been observed experimentally in, e.g., spatially extended nonlinear dissipative systems \cite{Purwins2010}, vertical-cavity semiconductor optical amplifiers \cite{Barbay2008}, nematic liquid crystal layers with {a spatially} modulated input beam \cite{Haudin2011}, and magnetic fluids \cite{Lloyd2015}.

	Homoclinic snaking in continuous systems was first described in \cite{Pomeau1986,Bensimon1988} to be caused by a pinning effect, by which the front locks to the pattern, resulting in a finite range of parameter values around the Maxwell
	point where a stationary localized solution can exist. The interval in which a snaking occurs is also therefore referred to as the pinning region that has been studied numerically in, e.g., \cite{Sakaguchi1996,Woods2006,Burke2006,Burke2007,Burke2007a}. In general, the pinning effect cannot be described by multiple asymptotics \cite{Pomeau1986}, i.e.,\ the length of the pinning region is exponentially small in a parameter which is related to the pattern amplitude. The approximation of the pinning region was provided by Kozyreff and Chapman \cite{Chapman2009,Kozyreff2006} and Dean \emph{et al.}\  \cite{Dean2011} using a beyond-all-order asymptotics and by Susanto and Matthews \cite{Susanto2011,Matthews2011} using variational methods. 
	
	Homoclinic snaking is also observed in spatially discrete systems, such as in the discrete bistable nonlinear Schr\"odinger equation \cite{Carretero-Gonzalez2006,Chong2009,Chong2011}, which leads to a subcritical Allen-Cahn
	equation \cite{Taylor2010}, optical cavity solitons \cite{Yulin2008,Yulin2010}, discrete systems with a weakly broken pitchfork bifurcation \cite{Clerc2011}, {and} in patterns on networks appearing due to Turing instabilities \cite{mccu16}. If in the continuous case the front locking is due to pattern formation, in the discrete systems it is due to the imposed lattice. The pinning region in this case was approximated analytically by Matthews and Susanto \cite{Matthews2011} and Dean \emph{et al.}\ \cite{Dean2015}. 
	
	Note that in all the aforementioned references, homoclinic snaking is studied either in continuous systems or discrete ones that no longer admit snaking in the continuum limit.  The transition of snaking structures from the discrete to the continuous limit is unfortunately rather lacking, which is particularly important because, e.g., when solving a continuous equation numerically, unavoidably one actually solves its discrete approximation. It is then necessary to recognize features that appear due to the discretization.  Here, we provide a comprehensive study on the subject. We consider the discrete cubic-quintic Swift-Hohenberg equation, obtained from discretizing the spatial derivatives of the (continuous) Swift-Hohenberg equation with central finite differences. 
	To our best knowledge, previous works on the discrete equation are only Peletier and Rodr\'iguez \cite{Peletier2004,Rodriguez2004}, who studied pattern formations in the system with a few {sites} only, and Collet \cite{Collet1998} that views the system as a discrete-time lattice map and analyzes the instability of homogeneous stationary solutions.

	Here, we report interesting and different properties that are not shared by the continuum  counterpart, such as multiple Maxwell points, i.e.,\ parameter values with periodic solutions having zero energies, bifurcation curves of periodic solutions exhibiting a snaking behavior, and localized states with complicated bifurcation diagrams. In general, we characterize three different regions of the discretization parameter, wherein the discrete Swift-Hohenberg equation behaves either similarly or differently from the continuum limit. Moreover, we provide theoretical analysis of the snaking and the pinning region {in the uncoupled limit}, i.e.,\ weak coupling region, through formal perturbation expansions, which is generally applicable to any strongly discrete system.
	
	The paper is outlined as {follows}. The spatially discrete Swift-Hohenberg equation is discussed in Sec. \ref{sec2}. In the section, we also study the stability of the uniform solutions. We discuss periodic solutions in Sec. \ref{sec3}. Section \ref{sec4} is on localized solutions and their asymptotic expressions that {are} obtained through multiple scale expansions. The width of the pinning region for varying parameters is also discussed in the section. We then derive {this} width asymptotically in the uncoupled limit in Sec. \ref{aap}, which {is} then compared with computational results, where good agreement is obtained. Conclusions are in Sec. \ref{sec6}.
	
	\section{Governing equation and {uniform} solutions}
	\label{sec2}
	
	The cubic-quintic Swift-Hohenberg equation is given by \cite{Budd2005}
	\begin{eqnarray}
	{\frac{\partial u}{\partial t}} &=& r u - \left(1+{\frac{\partial^2 }{\partial x^2}}\right)^2u + b_3u^3-b_5u^5,
	\label{eq:Swift-Hohenberg}
	\end{eqnarray}
	where $u = u(x, t)$ is {a} scalar function defined on the {real} line, $r$ is a real bifurcation parameter (control or stress parameter) \cite{Getling1998}, and $b_3$ and $b_5$ are nonlinearity coefficients.
	Equation \eqref{eq:Swift-Hohenberg} is invariant under $x\rightarrow -x$ and $u \rightarrow -u$. Without loss of generality, by scaling one can take parameter $b_5=1$ 
	\cite{Getling1998,Burke2007a}. 
	
	The discrete Swift-Hohenberg equation is obtained from \eqref{eq:Swift-Hohenberg} by discretizing the spatial derivatives using {central} finite difference
	\begin{eqnarray}
	\frac{d u_n}{d t}&=& (r-1)u_n-\frac{2}{{h^2}}\Delta_2 u_n - \frac{1}{h^4}\Delta_4 u_n\nonumber\\
	&&+b_3u_n^3-b_5u_n^5,\label{eq:dSwift-Hohenberg}
	\end{eqnarray}
	where $\Delta_2 u_n={u_{n+1}-2u_n+u_{n-1}},$ $\Delta_4 u_n={u_{n+2}-4u_{n+1}+6u_n-4u_{n-1}+u_{n-2}},$ and  $h$ is the discretization parameter. In the results presented below, mostly we take $b_3=2$. However, we also consider different values of the parameter. 
	
	In this work, we study the time-independent solution of Eq. \eqref{eq:dSwift-Hohenberg}, i.e.,\
	\begin{eqnarray}
	\frac{du_n}{dt}=0.\label{eq:dSwift-Hohenberg_ti}
	\end{eqnarray}
	Equation \eqref{eq:Swift-Hohenberg} can be written as $\frac{d u_n}{d t}=-P\frac{\delta {E}}{\delta u_n}$, where the Lyapunov function $E$, referred to as the energy function of the system, is given by
	\begin{eqnarray}
	\displaystyle E&=&\frac1P\sum_{n=1}^{P}\left\{-\frac{1}{2}(r-1){u_{{n}}}^{2}\nonumber\right.\\
	&&\left.-\frac{1}{2}\left( {\frac {  \left( u_{{n
					+1}}-u_{{n}} \right) ^{2}+ \left( u_{{n}}-u_{{n-1}} \right) ^{2}
		}{{h}^{2}}}\right)\nonumber\right.\\
	&&\left.+\frac{1}{2}\,{\frac { \left( u_{{n-1}}-2\,u_{{n}}+u_{{n+1}} \right) ^{2}}{{h}^{4}}}-\frac{1}{4}\,b_{{3}}{u_{{
				n}}}^{4}+\frac{1}{6}\,b_{{5}}{u_{{n}}}^{6}\right\},\nonumber\\
	\label{eq:Ly}
	\end{eqnarray}
	and $P$ is the period of {the solution, i.e.,\ $u_{n+P}=u_n$}.
	
	The discrete Swift-Hohenberg equation {has} the same uniform solution $u_n=U_j$ as the continuum limit studied in \cite{Burke2007a}, which is given by
	\begin{eqnarray}
	0=(r-1)U_j+b_3 U_j^3-b_5 U_j^5,\label{eq:unisol}
	\end{eqnarray}
	that can be solved to yield
	\begin{eqnarray}
	U_0 = 0,\,U_+= \left[\frac{1}{2b_5}\left(b_3\pm \sqrt{b_3^2+4b_5\left(r-1\right)}\right)\right]^{\frac{1}{2}}, \label{eq:unisol_u}
	\end{eqnarray}
	and its mirror symmetric $U_- = -U_+$. The bifurcation diagram of the uniform solutions is shown in Fig.~\ref{fig:unisol_hg2}.
	The two branches  of $U_+$ collide at 
	\[r_1=1-\frac{b_3^2}{4b_5}\] and $U_+$ with the minus sign bifurcates from $U_0$ at $r_2=1$.

	To determine the linear stability of a solution ${\tilde{u}_n}$, we write
	\begin{eqnarray}
	u_n =\tilde{u}_n+\epsilon e^{\lambda t}\hat{u}_n.
	\label{eq:anz_d}
	\end{eqnarray}
	After substituting the ansatz into equation \eqref{eq:dSwift-Hohenberg} and linearizing it about
	$\epsilon=0$, we obtain the linear equation 
	\begin{equation}
	\lambda \hat{u}_n = \mathcal{L} \hat{u}_n,
	\label{evp}
	\end{equation}
	where $$\mathcal{L}:= r-1-\frac2{h^2}\Delta_2-\frac1{h^4}\Delta_4+3b_3\tilde{u}_n^2-5b_5\tilde{u}_n^4 $$
and {the spectrum } $\lambda$ {defines the stability} of the solution $\tilde{u}_n$. A solution is said to be stable when all {$\lambda\leq 0$} and unstable when $\exists \, \lambda>0$. The spectrum of the linear differential operator $\mathcal{L}$ on the infinite dimensional space is the set of all complex numbers $\lambda$ such that $(\mathcal{L}-\lambda)$ either has no inverse or is unbounded. In general, the spectrum of the linear operator will consist of {a} continuous spectrum and {a} discrete spectrum (eigenvalue) \cite{hunt11}.

For the uniform solution $\tilde{u}_n=U_j$, $j=0,+,-$, one has $\hat{u}_n=e^{ikhn}$, where $k$ is the wavenumber of the perturbation, from which we obtain the dispersion relation
	\begin{eqnarray}
	{\lambda\left(k\right)}&=&r-1+3\,b_{{3}}{U_{{j}}}^{2}-5\,b_{{5}}{U_{{j}}}^{4}\nonumber\\
	&&-4\left({\frac {\cos \left( kh \right)-1 }{{h}^{2}}}\right)\left(1+{\frac {\cos \left( kh \right) -1}{{h}^{2}}}\right).\label{eq:eig_fun}
	\end{eqnarray}
The continuous spectrum is the interval of values that can be attained by $\lambda$ for all $k\in\mathbb{R}$. The point $r_0$, i.e.,\ $j=0$ in \eqref{eq:eig_fun}, corresponds to the condition when the maximum of the function touches the horizontal axis, which is attained at the wave number
	\begin{eqnarray}
	{k =  \frac{1}{h}\left({ {\pi \pm \arccos \left( \frac{1}{2} h^2-1 \right) }}\right),}
	\label{eq:khl2}
	\end{eqnarray}
	for $h< 2$ and
	\begin{eqnarray}
	{k=\pm\left(\frac{\pi}{h}\right),}
	\label{eq:khg2}
	\end{eqnarray}
	for $h\geq2$. The numbers are important in the study of bifurcating periodic solutions and localized solutions below. They will be the wave numbers of the carrier wave of the localized solutions. 

	\subsection{Stability for $h < 2$}
	By substituting Eq.\ \eqref{eq:khl2} into \eqref{eq:eig_fun} and considering $j=0$ and $\lambda\left(k\right)=0$, we obtain that $U_0$ changes stability at
	\begin{eqnarray}
	r_0 = 0
	.
	\label{eq:r0_hl2}
	\end{eqnarray}
	Using the same procedure for $U_{+}$, we obtain that it changes stability at
	\begin{flalign}
	r_{+}=\frac{5}{4}-\frac{b_3}{8b_5}\left(b_3+\sqrt{b_3^2+4b_5}\right).
	\label{eq:rpm_hl2}
	\end{flalign}
	Equations \eqref{eq:r0_hl2} and \eqref{eq:rpm_hl2} indicate that the stability of $U_0$ and $U_{+}$ does not depend on the discretization parameter for $h<2$. The stability of the uniform solutions is depicted in Fig.~\ref{fig:unisol1}, which is the same as the continuous Swift-Hohenberg equation \cite{Burke2007a}.
	
	\subsection{Stability for $h \geq 2$}
	
	Following the same steps as the case of $h<2$, we obtain that for $h\geq2$ the stability change for $U_0$ and $U_{+}$ {occurs}, respectively, at
	\begin{align}
		r_0=&1-{\frac {8}{{h}^{2}}}+\frac{16}{h^4},
	\label{eq:r0_hg2}\\
	r_{+}=&\left(1+\frac{2}{h^2}-\frac{4}{h^4}\right)\nonumber\\
	&-\frac{b_3}{8b_5}\left(b_3+\frac{1}{h^2}\sqrt{h^4\,b_3^2+32b_5\left(h^2-2\right)}\right).
	\label{eq:rpm_hg2}
	\end{align}
	The stability of the uniform solutions now depends on $h$. 
	
	\begin{figure}[t!]
		\subfigure[]{\includegraphics[scale=0.5]{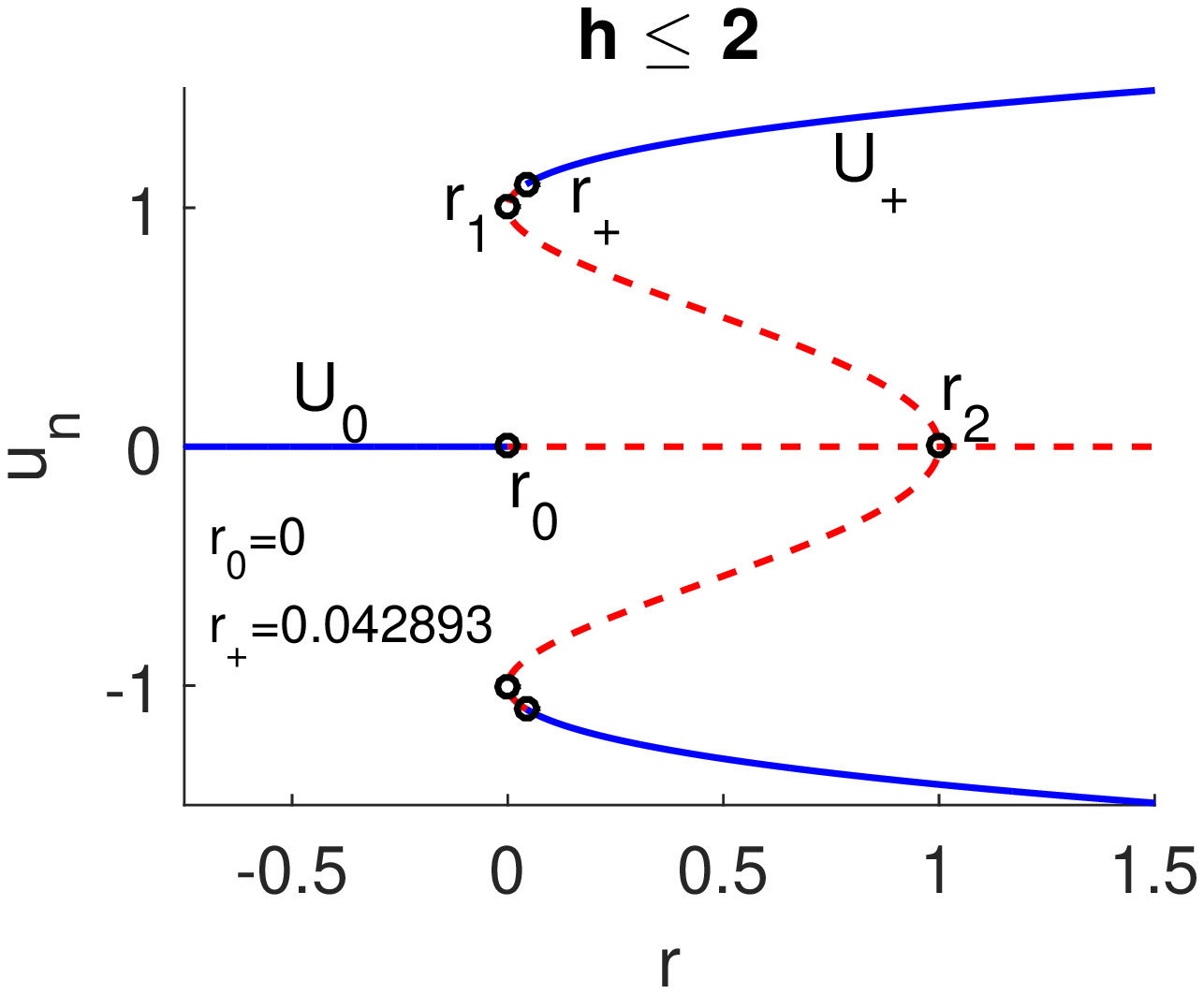}\label{fig:unisol1}}
		\subfigure[]{\includegraphics[scale=0.5]{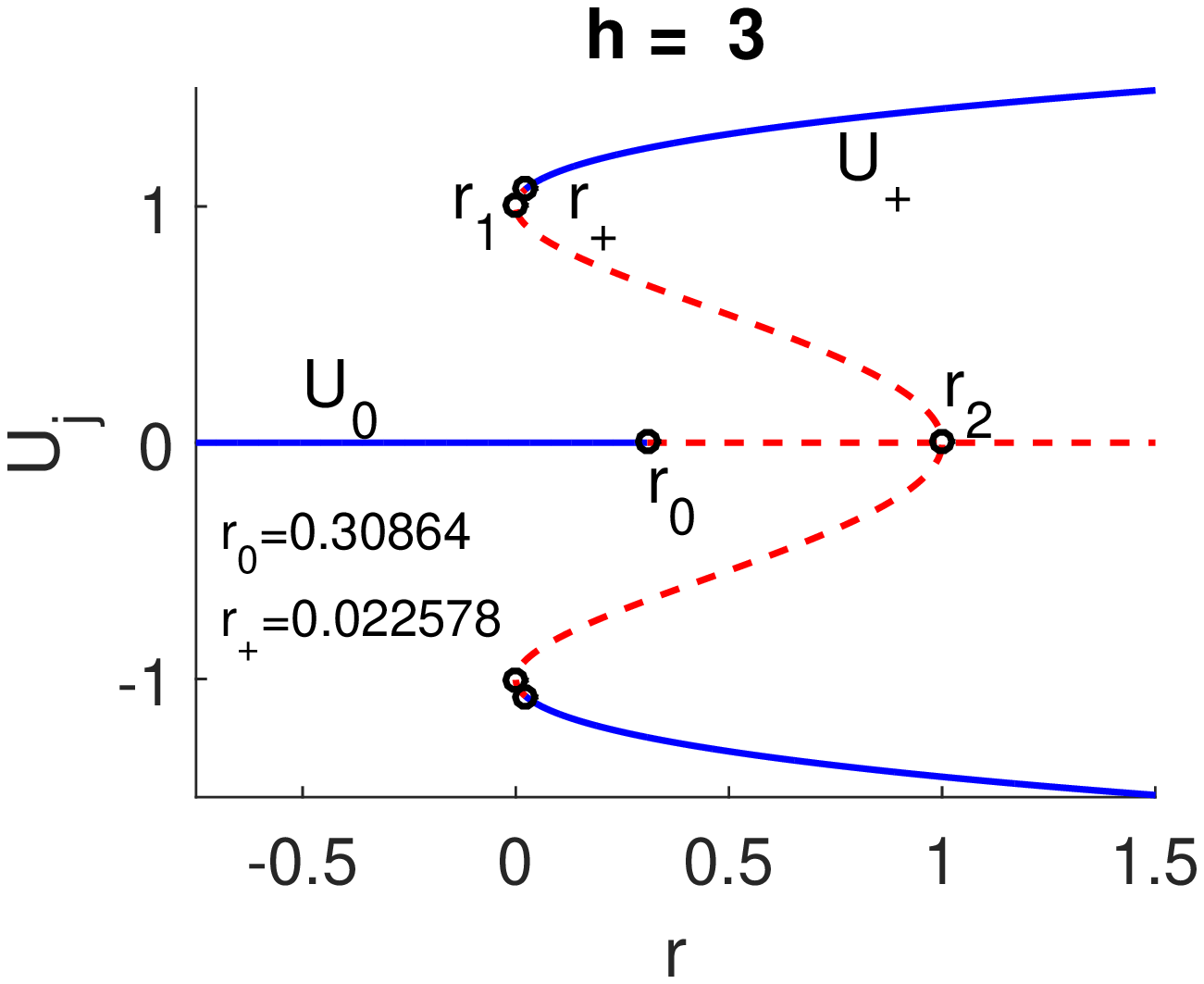}\label{fig:unisol2}}
		\caption{The existence curve of the uniform solutions of the governing equation \eqref{eq:dSwift-Hohenberg}.
			 {Blue solid and red dashed lines indicate{, respectively,} stable and unstable solutions.}}
		\label{fig:unisol_hg2}
	\end{figure}
	
	Figure \ref{fig:unisol2} shows the bifurcation diagram of the uniform solutions for $h=3$.
	The point $r_0$ at which $U_0$ changes its stability is shifted to the right.
	In the limit $h\rightarrow\infty$, the stability of $U_0$ changes at $r_0=1$.
	The stability of $U_{+}$ also changes as a function of $h$.
	We can see that $r_{+}$ is getting closer to $r_1$ as $h$ increases and in the limit when $h\rightarrow\infty$, the stability of $U_{+}$ changes at $r_{+}=r_1=0$.
	
	One main difference between the uniform solutions of the continuous and the discrete equations is that in the strongly discrete case ($h>2$), one can have a bistability between $U_0$ and $U_+$, i.e.,\ when $r_0>r_+$.
	
	\begin{figure}[t!]
		\centering\
		\includegraphics[scale=0.56]{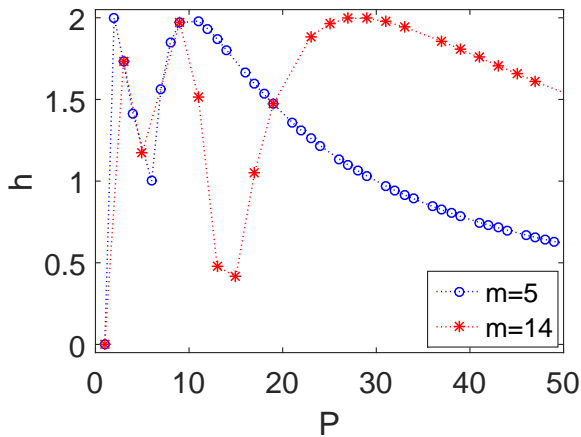}
		\caption{The relation between the discretization parameter $h$ and the period $P$ for {$h\leq 2$ and two values of $m$, i.e.,\ $m=5$ and $m=14$}.
		}
		\label{fig:P_vs_h}
	\end{figure}
		
	\section{Periodic solutions}
	\label{sec3}
	
	The discrete Swift-Hohenberg equation also admits periodic solutions that bifurcate from the uniform solution $U_0$ at $r=r_0$. We can obtain an approximation to the bifurcating periodic solution by writing
	\begin{eqnarray}
	{u_{P,n}} &=&U_{0}+\hat{\varepsilon}\,\cos\left(khn\right),
	\label{eq:perturb}
	\end{eqnarray}
	with $\hat{\varepsilon}$ small and {$k$} given by \eqref{eq:khl2} or \eqref{eq:khg2}. 
	Note that for the continuous function $f(x)=\cos(kx)$, its period is easily given by $P=2\pi/k$. For the discrete function $f_n=\cos(khn)$, the period is calculated differently \cite{Oppenheim1983}, i.e.,\ it is periodic with period $P\in \mathbb{Z^+}$ if $\exists\,{m} \in \mathbb{Z^+}$ {that does not have any factor in common with $P$}, such that
	\begin{eqnarray}
	P=  \frac{2\pi m}{k\,h}. 
	\label{eq:kh_d}
	\end{eqnarray}
	The solution \eqref{eq:perturb} is therefore periodic only if there are integers $m$ and $P$ with no common factors that satisfy {\eqref{eq:kh_d}}. For $h<2$, using \eqref{eq:khl2} the plot of \eqref{eq:kh_d} is shown in Fig.\ \ref{fig:P_vs_h} , relating the discretization parameter $h$ and the period $P$ for several values of $m$.
	Note that not every $h<2$ will yield periodic solutions. There are values of the parameter that correspond to almost-periodic (i.e.,\ {quasi-periodic}) functions. However, {the study of these quasi-periodic solutions} is beyond the scope of the present paper and is addressed for future work. For $h\geq2$, \eqref{eq:kh_d} with \eqref{eq:khg2} implies that all the bifurcating periodic solutions have period $P=2$.

	Substituting Eq.\ \eqref{eq:perturb} into the energy function \eqref{eq:Ly} and finding the minimum of $E$, i.e.,\
	\begin{eqnarray}
	\frac{\partial E}{\partial\hat{\varepsilon}}=0,
	\end{eqnarray}
	{yield} an approximate amplitude $\hat{\varepsilon}$ of the periodic solutions about $U_0$, that is given by
	\begin{flalign}
		\hat{\varepsilon}(r) =\left( \frac{b_3-(4rb_5+b_3^2)^{\frac{1}{2}}}{2b_5}\right)^{\frac{1}{2}}
		\label{eq:A_hl2}
	\end{flalign}
	for $h<2$ and
	\begin{flalign}
		\hat{\varepsilon}(r)=\left(\dfrac{h^2b_3-\left[\left(b_5\left(4r-1\right)+b_3^2\right)h^4+32b_5\left(h^2-2\right)\right]^{\frac{1}{2}}}{2b_5}\right)^{\frac{1}{2}}
		\label{eq:A_hg2}
	\end{flalign}
	for $h\geq2$. One can also perform asymptotic analysis using multiple scale expansions to obtain the bifurcating periodic solution. This is presented in Appendix [see (\ref{eq:Un_zero})].

	We solve Eq.\ \eqref{eq:dSwift-Hohenberg_ti} numerically using a Newton-Raphson method with periodic boundary conditions and using \eqref{eq:perturb} and \eqref{eq:A_hl2} or \eqref{eq:A_hg2} as an initial guess in our numerics. Note that, herein, we take the computational number of sites $N$ to be a multiple of $P$.
	We use a pseudo-arclength method to continue the computations past limit points \cite{Seydel1994}.
	We present the bifurcation diagram in the $\left(r,||u||\right)$ plane with
	\begin{eqnarray}
	||u||= {\left(\frac{1}{N}\sum^N_{n=1}{u_n^2}\right)^\frac{1}{2}}.\label{eq:norm_d}
	\end{eqnarray}
	
	After a periodic solution is found, we determine its stability by solving the eigenvalue problem \eqref{evp}, where $\tilde{u}_n$ is now a periodic solution, i.e.,\ $\tilde{u}_n=u_{P,n}$. {At the same time, we also seek for its Maxwell points $r_{M1}$, i.e.,\ points where the periodic state $u_{P,n}$ and the zero solution $U_0$ have the same energy ($E[u_{P,n}]=E[U_0]=0$).}
	
	In the next subsections, we divide the parameter interval into three regions, i.e.,\ $h<1$, $1\leq h<2$, and $h\geq 2$. The main reason is the qualitative features of the solutions in each region, which are distinguishably different.

	\subsection{Periodic {solutions} for $h<1$}
	
	\begin{figure}[htbp!]
		\centering
		{\includegraphics[scale=0.56]{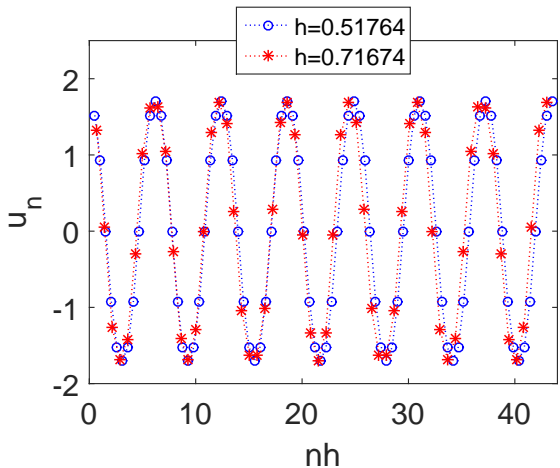}
		\caption{{Periodic solutions for $h=0.5176$ and $0.7167$ {for} $r=1$.} 
			}
		\label{subfig:U_m5m7}}
	\end{figure}	
	\begin{figure*}
		\centering
		\subfigure[]{\includegraphics[scale=0.5]{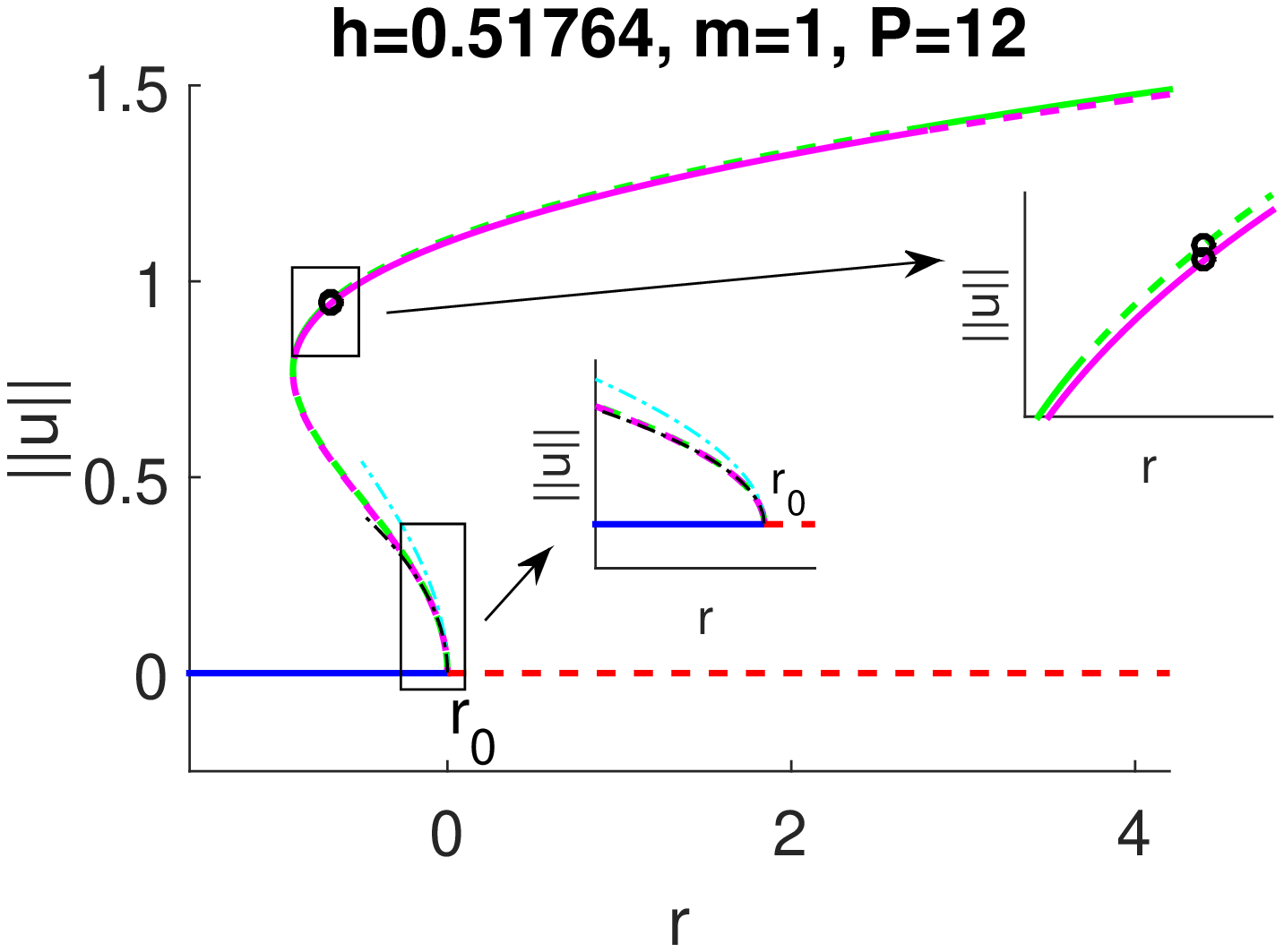}\label{subfig:persol_m5}}
		\subfigure[]{\includegraphics[scale=0.5]{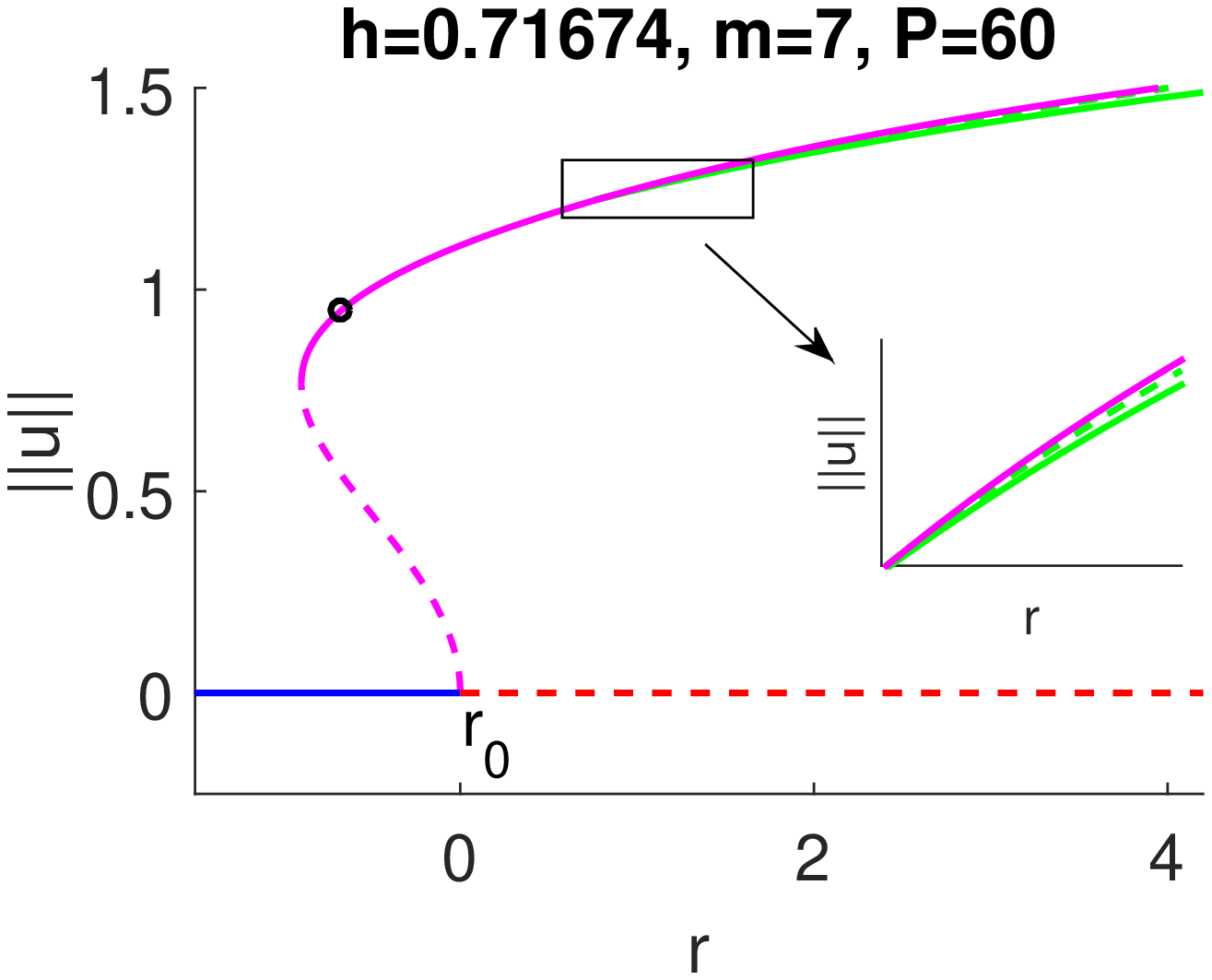}\label{subfig:persol_m7}}
		\subfigure[]{\includegraphics[scale=0.5]{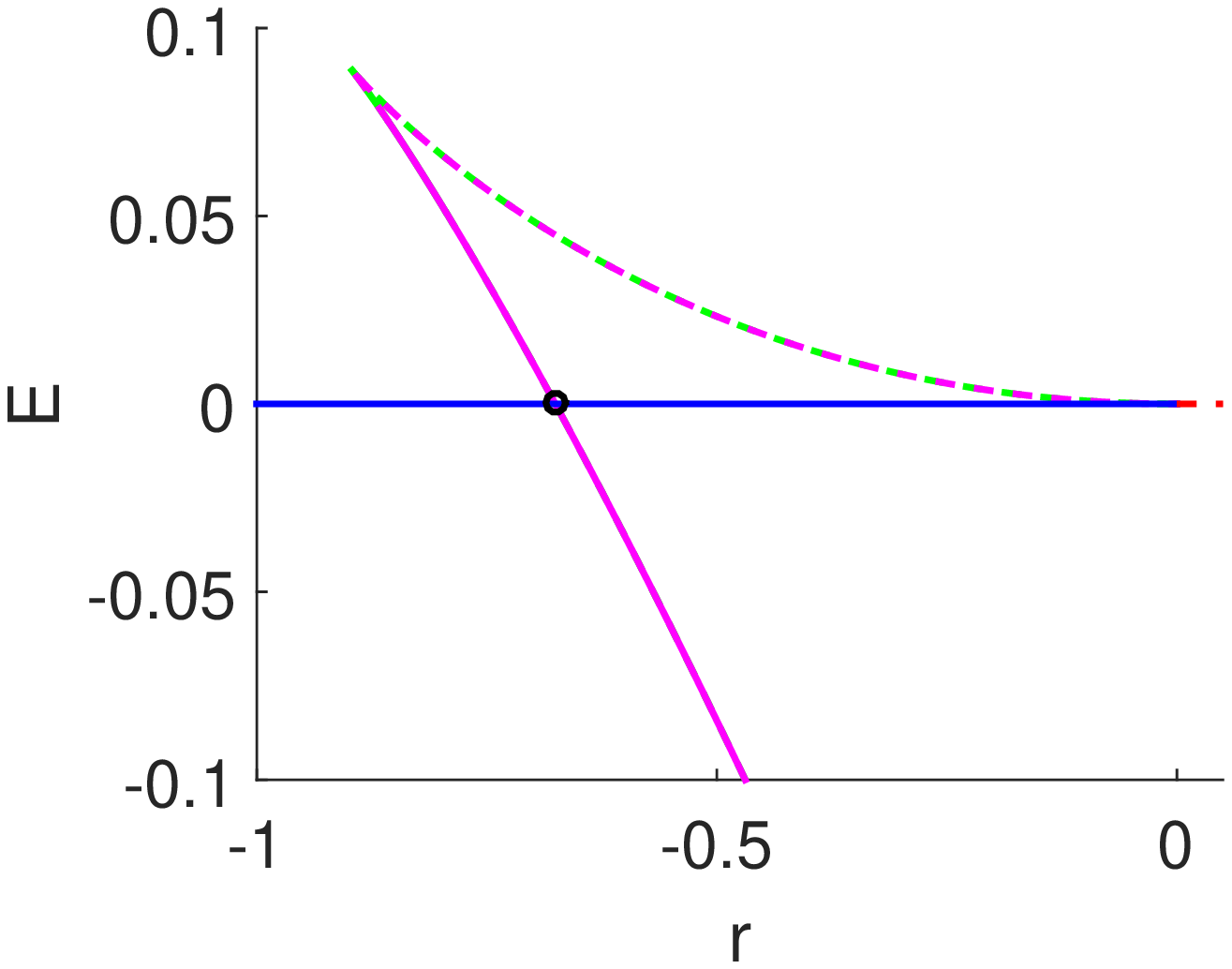}\label{subfig:Ly_m5}}
		\subfigure[]{\includegraphics[scale=0.5]{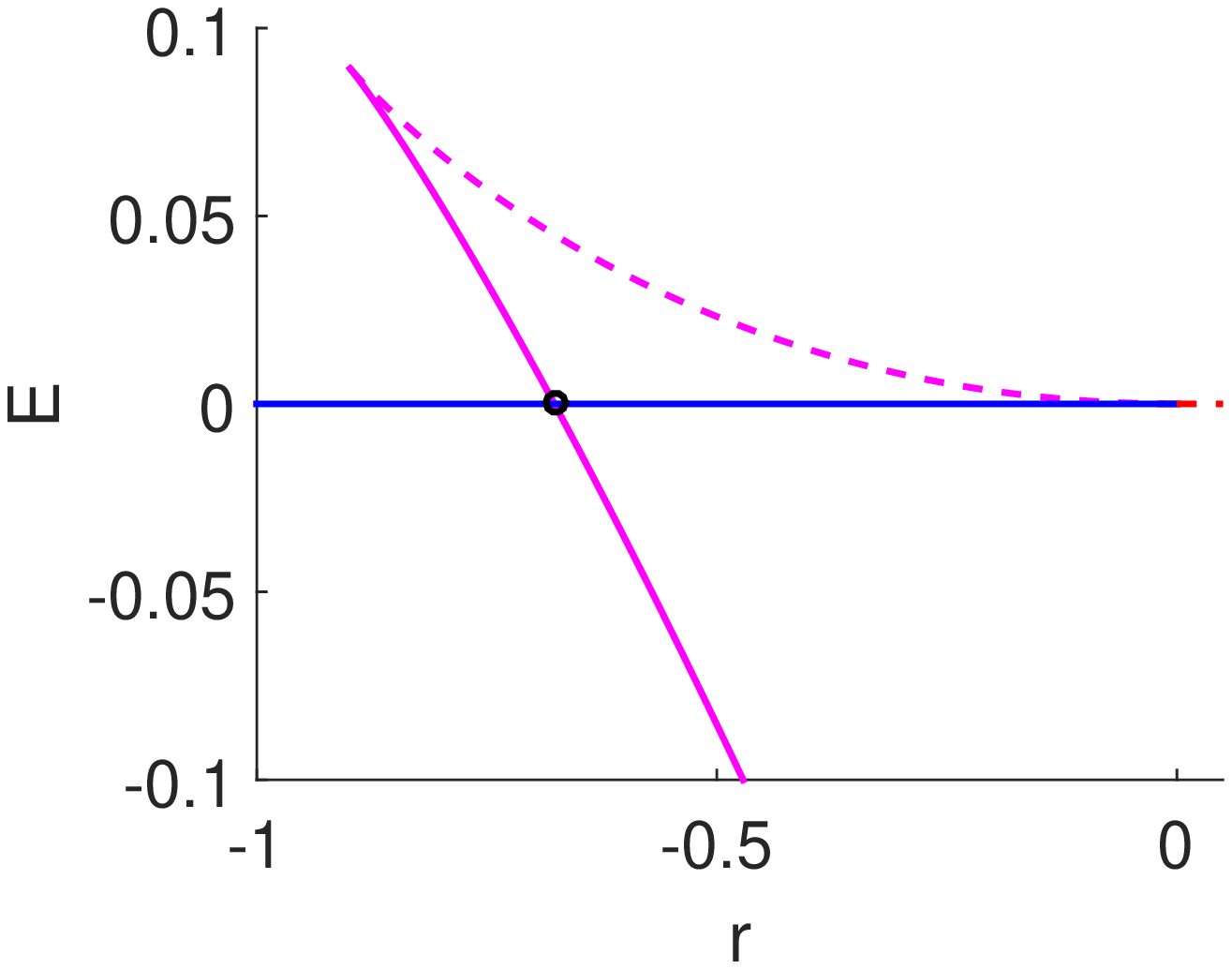}\label{subfig:Ly_m7}}	
		\subfigure[]{\includegraphics[scale=0.5]{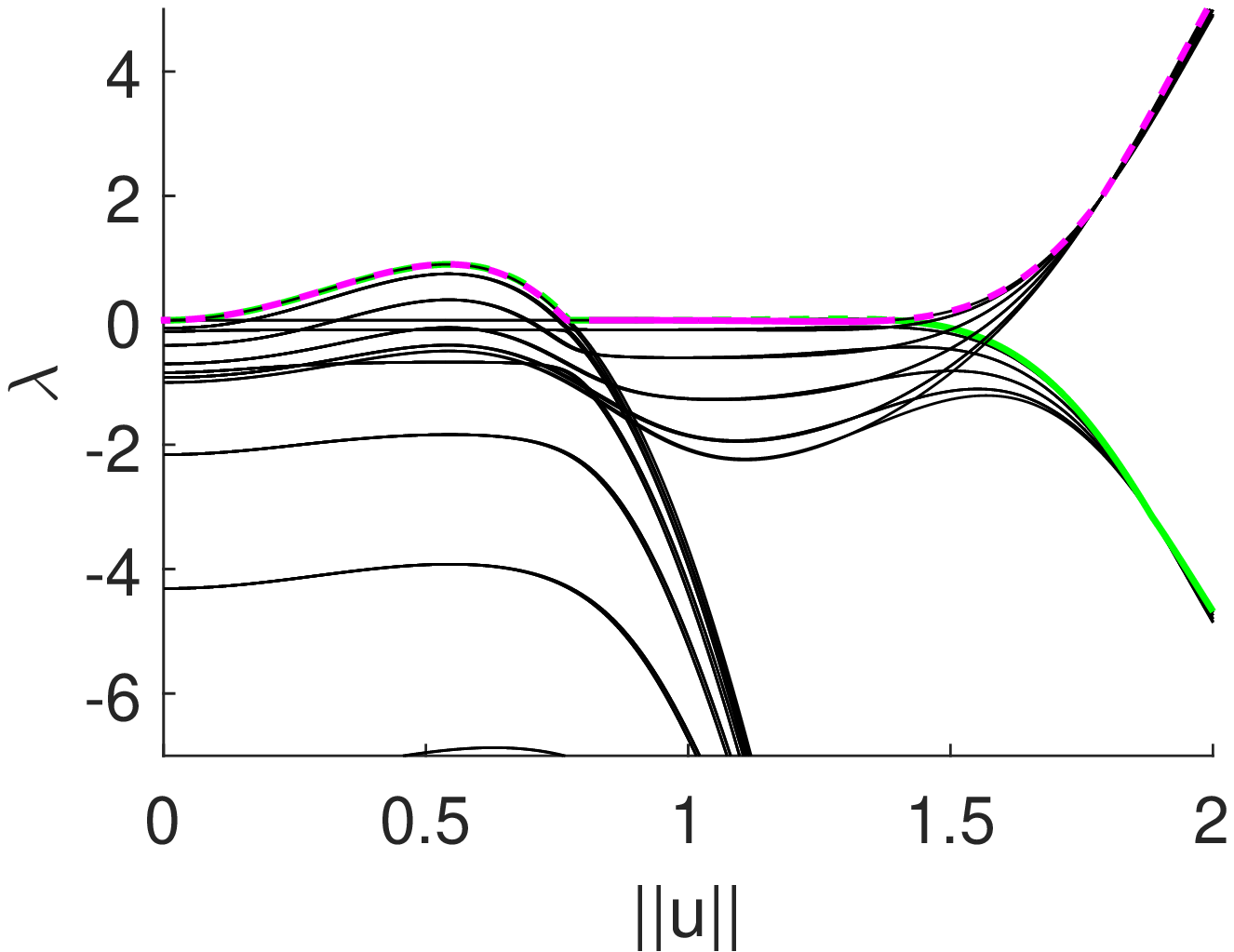}\label{subfig:eigps_m5}}		
		\subfigure[]{\includegraphics[scale=0.5]{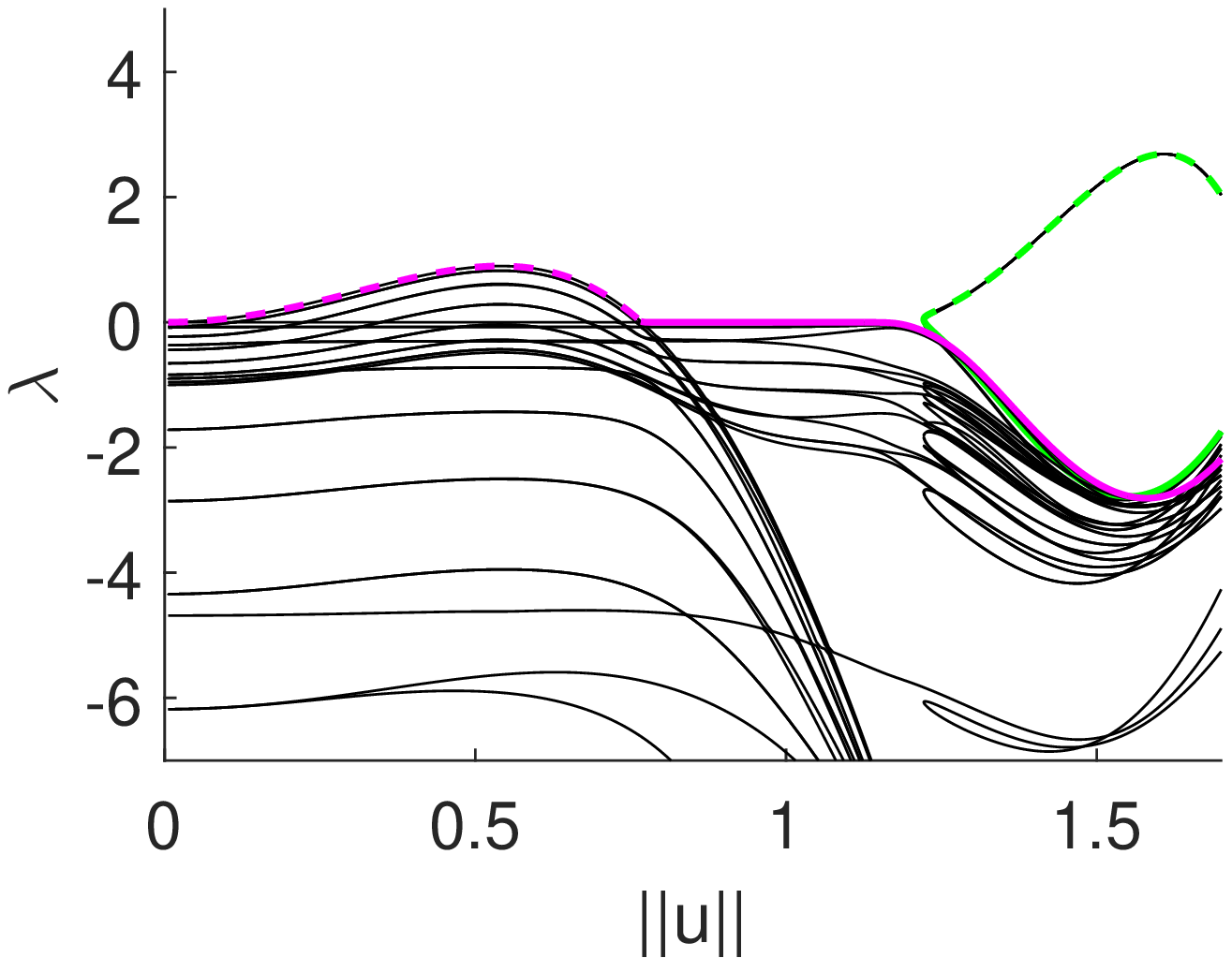}\label{subfig:eigps_m7}}			
		\caption{ {(a), (b) Bifurcation diagrams of periodic solutions for (a) $h=0.5176$ and (b) $h=0.7167$.
				(c), (d) The energy function of the periodic solutions. 
				(e), (f) Eigenvalues of the periodic solutions.			
				The magenta and green lines correspond to periodic solutions along the main and extra bifurcation curves, respectively.
				Black thin lines in (e) and (f) indicate non-critical eigenvalues of the periodic solutions.
				Solid and dashed lines correspond to stability and instability, respectively.
				Circles indicate Maxwell points.} The dashed-dotted cyan and black lines in panel (a) are amplitudes \eqref{eq:perturb} and \eqref{eq:Un_zero}, respectively.				
		}
		\label{fig:persol_m5m7}
	\end{figure*}	

	Figure \ref{subfig:U_m5m7} shows the profile of two periodic solutions for $h=0.5176$ and $0.7167$ which correspond to $P=12, m=1$ and $P=60, m=7$, i.e.,\ the second solution has a period of five times larger than the first.	
	We choose {these} two values of $h$ that are representative to the case $h<1$.
	
	{Figures}\ \ref{subfig:persol_m5} and~\ref{subfig:persol_m7} show the bifurcation diagram and the stability of the periodic solutions for the two values of $h$ above. The diagrams are similar to those of the continuous Swift-Hohenberg equation \cite{Burke2007a}. 
	However, the discretization causes the appearance of an additional branch and possibly a Maxwell point. Note that in the continuous case, {periodic} solutions only have one upper branch and one Maxwell point \cite{Burke2006,Burke2007}. In panel (a), we also plot the analytical approximation  \eqref{eq:perturb} and \eqref{eq:Un_zero}, that is obtained using multiple scale asymptotics, showing good agreement. 
	
	\begin{figure*}[htbp!]
		\centering
		\subfigure[]{\includegraphics[scale=0.56]{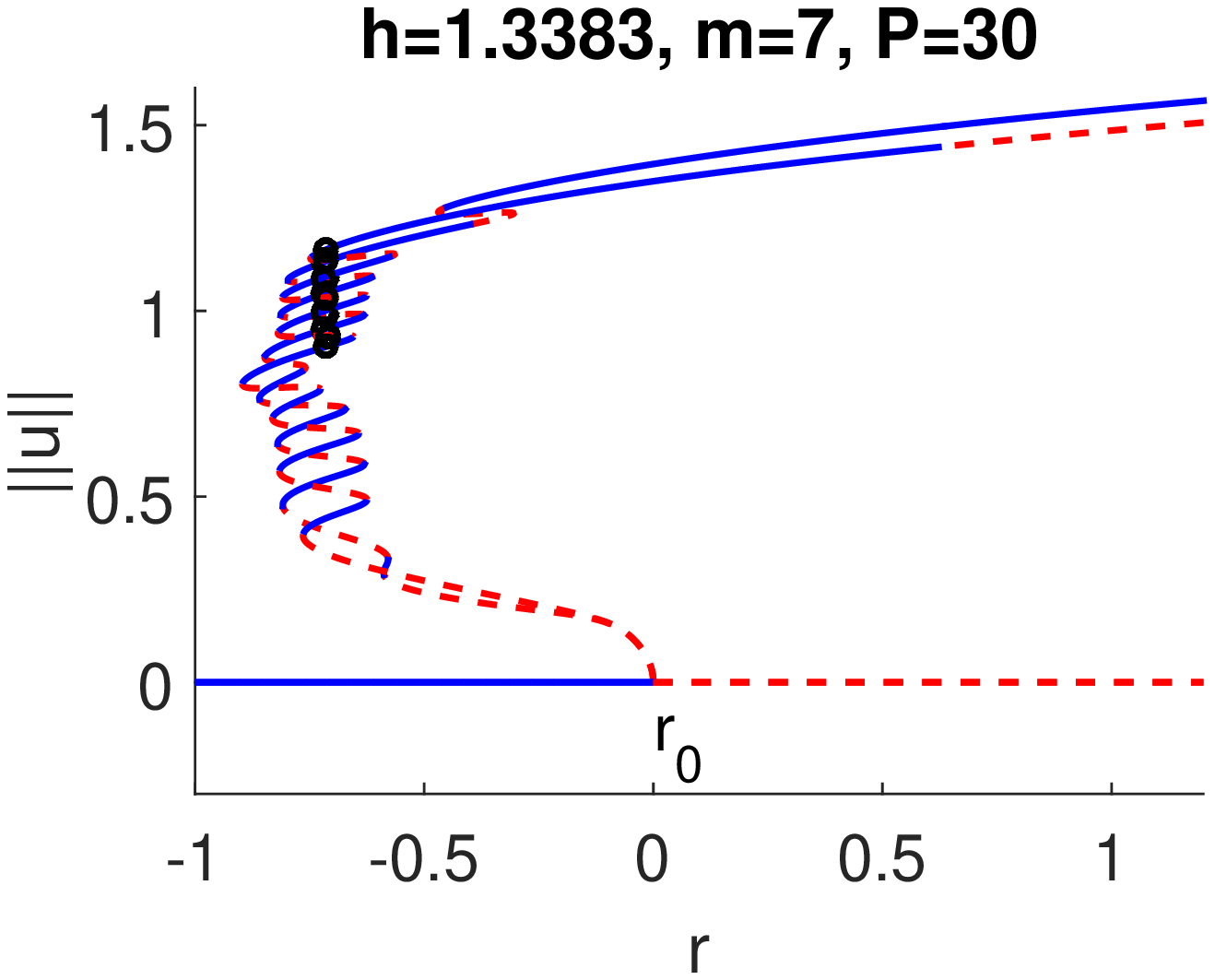}\label{subfig:persol_m14}}
		\subfigure[]{\includegraphics[scale=0.56]{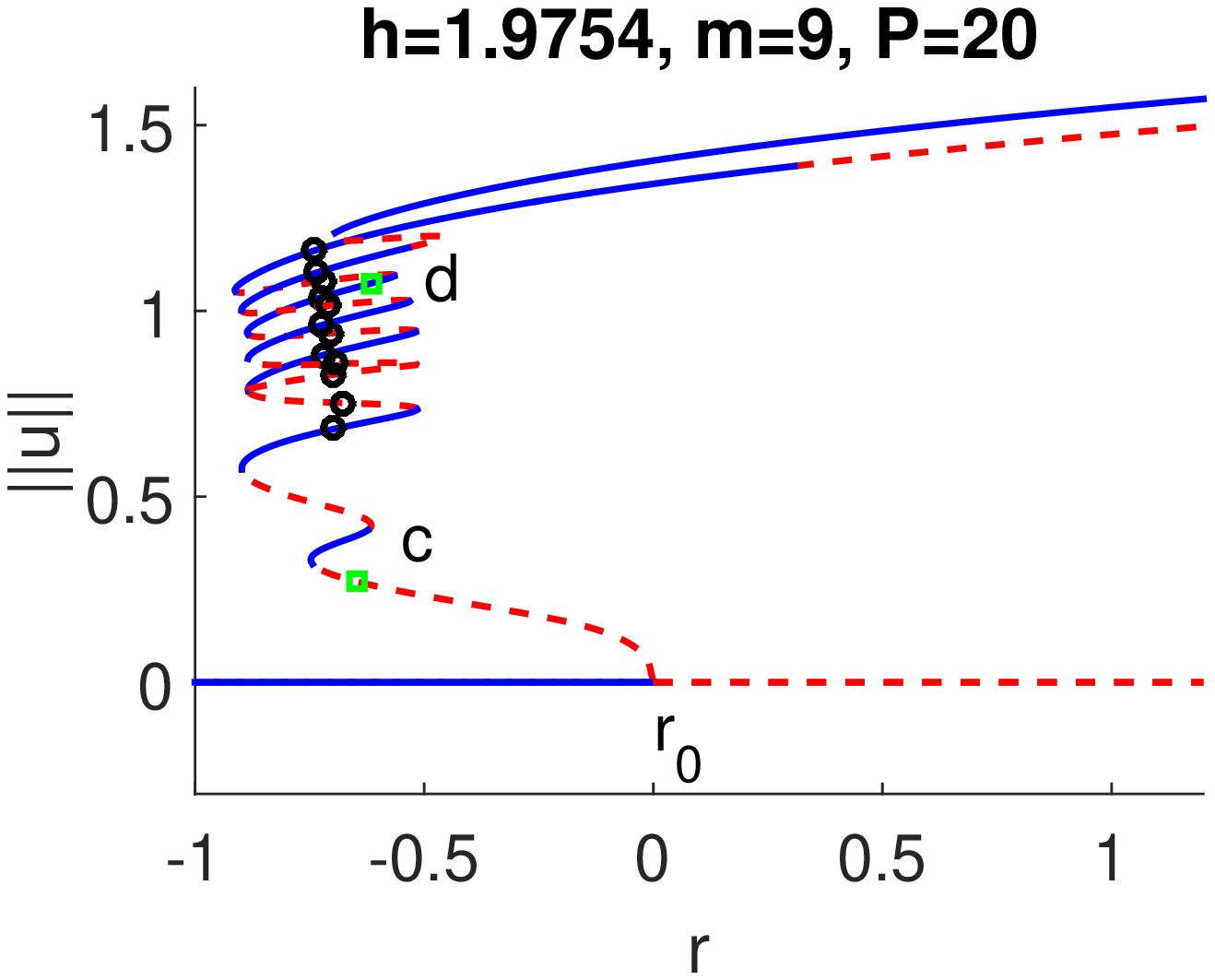}\label{subfig:persol_m27}}
		\subfigure[\,Solution profile at point c, $r=-0.6447$.]{\includegraphics[scale=0.56]{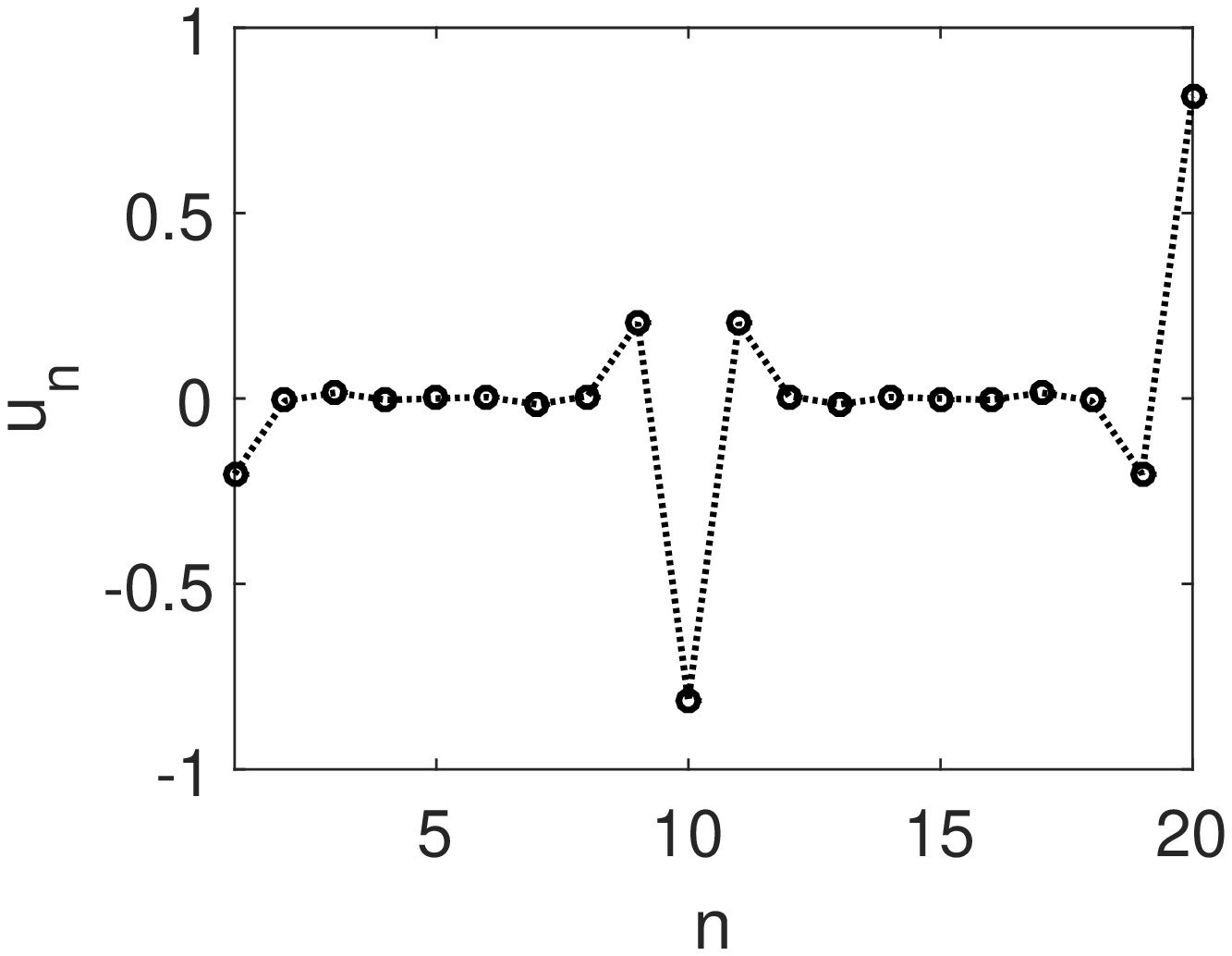}\label{subfig:persol_prom27a}}
		\subfigure[\,Solution profile at point d, $r=-0.6158$.]{\includegraphics[scale=0.56]{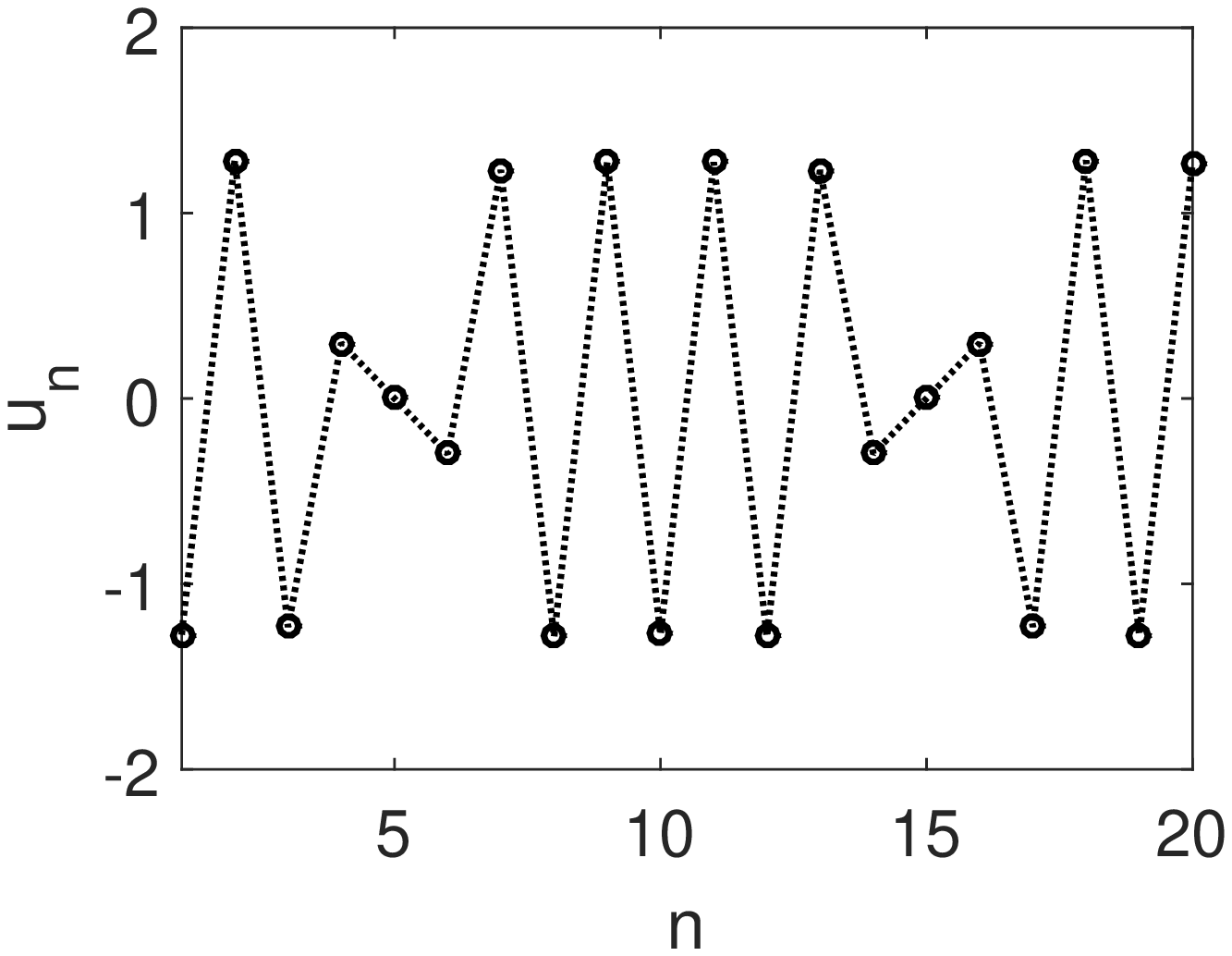}\label{subfig:persol_prom27c}}
		\caption{{(a), (b)} The bifurcation diagram of periodic solutions for two values of $h$ with $1\leq h <2$.
			{(c), (d)} Solution profiles for $h=1.9754$ at several values of $r$, indicated in (b).
			}
		\label{fig:persol_m14m25m26m27}
	\end{figure*}
	
	\begin{figure*}[htbp!]
		\centering
		\subfigure[]{\includegraphics[scale=0.56]{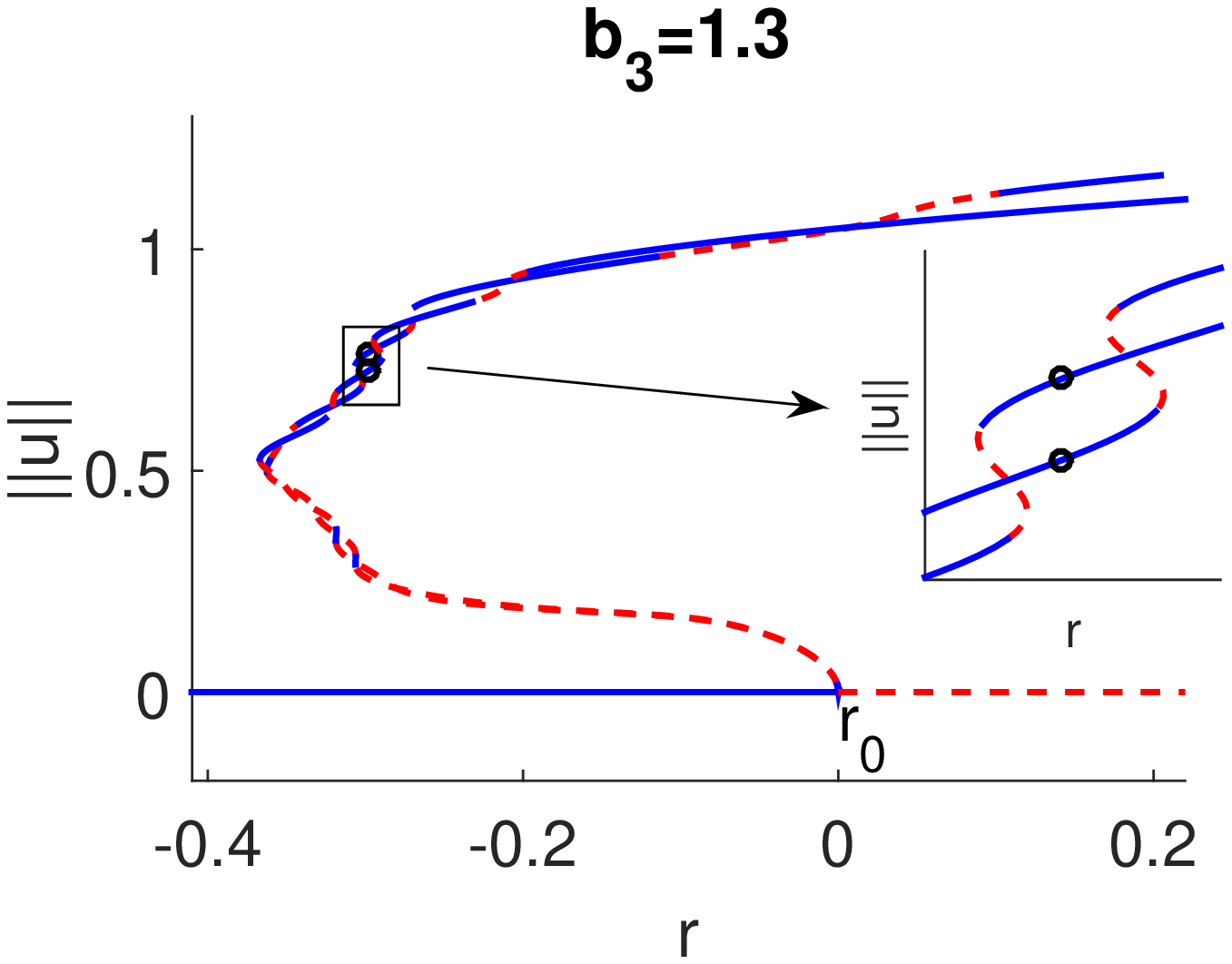}\label{subfig:persolm14b3_13}}
		\subfigure[]{\includegraphics[scale=0.56]{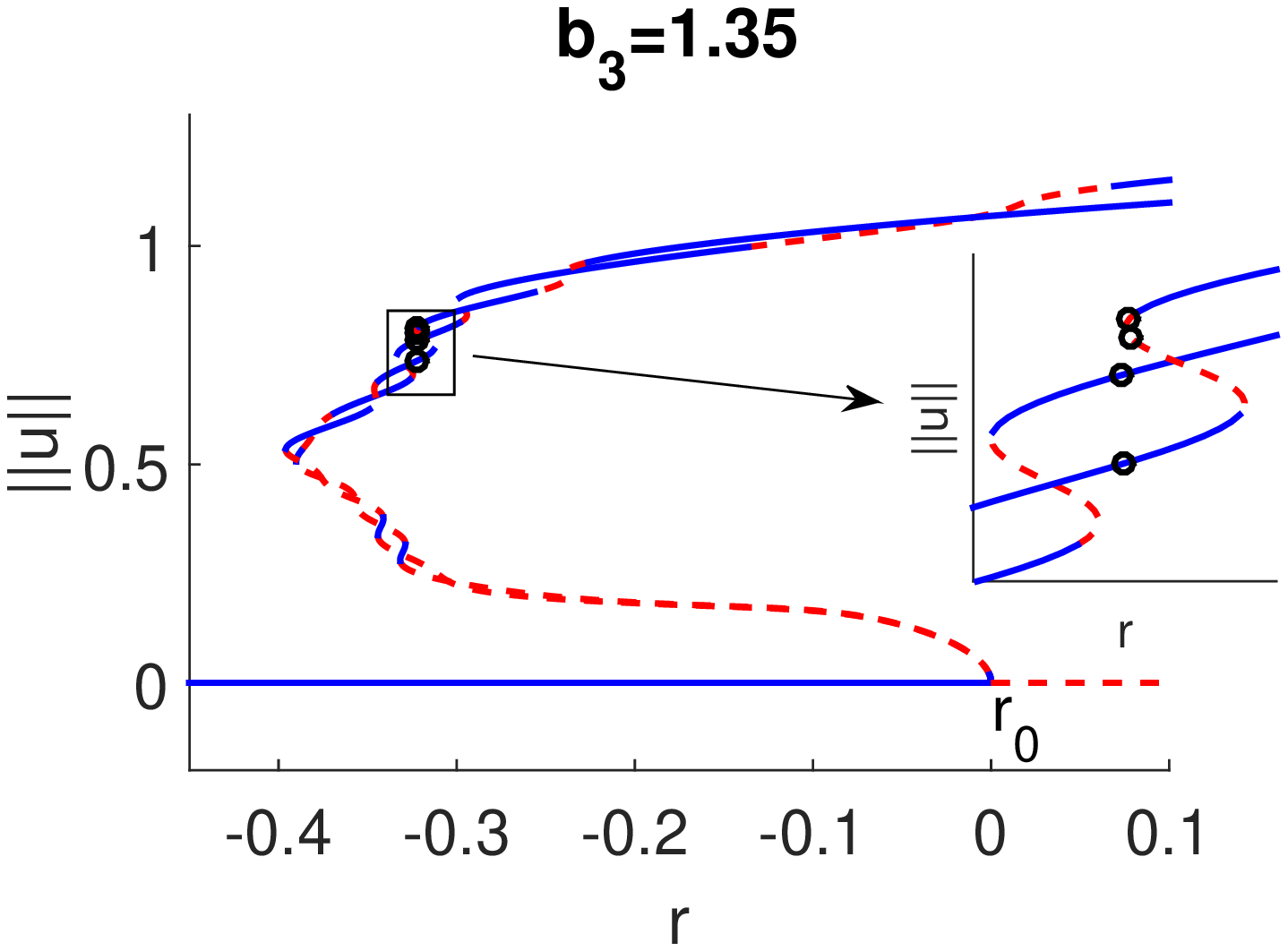}\label{subfig:persolm14b3_15}}
		\caption{The appearance of additional Maxwell points as we vary parameter $b_3$ for $h=1.3383$, $m=7$, and $P=30$.}
		\label{fig:persol_m14b3}
	\end{figure*}
	
	{Figures}\ \mbox{\ref{subfig:Ly_m5}} and \mbox{\ref{subfig:Ly_m7}} show the energy function of the periodic solutions for the two values of $h$ above. 
	Both panels show similar plots to those of the continuous Swift-Hohenberg equation. 
	The energy curves of the periodic solution $E[u_P]$ cross the horizontal axis at Maxwell points. In panel (c) the points are at $r_{M1}=-0.6755$ (stable) and $r_{M1}=-0.6754$ (unstable), while in panel (b) the (stable) Maxwell point is at $r_{M1}=-0.6762$. The free energy curves of the two upper branches are indistinguishably close to each other.

	{Figures}\ \mbox{\ref{subfig:eigps_m5}} and \mbox{\ref{subfig:eigps_m7}} show the spectrum of the periodic solutions along the two branches. Green and magenta lines indicate the critical eigenvalues of the periodic solutions along the primary and secondary upper branch.
	
	We also considered several other values of the discretization parameter $h$. The main difference between the continuous and the weakly discrete case is indeed the presence of an extra branch of periodic solutions that may also contain an additional Maxwell point. We conjecture that the splitting point where the primary and the secondary upper branches emerge moves to $r\to\infty$ as $h\to 0$, even though it may not increase uniformly. Note that in Fig.\ \ref{subfig:persol_m5} the value of $h$ is smaller than that in Fig.\ \ref{subfig:persol_m7}, but the branching point of the former occurs at a smaller value of $r$ than that of the latter. Additionally there can be changes of the stability of the periodic solutions along the upper branches.

	\subsection{Periodic {solutions} for $1\leq h<2$}	
	
	{Figures}\ \ref{subfig:persol_m14} and \ref{subfig:persol_m27} show the bifurcation diagrams of several periodic solutions for {two values of $h$ in the interval} $1\leq h<2$. The diagrams show snaking behavior with multiple Maxwell points along the stable and unstable branches, which was not seen in the previous case $h<1$ (including the continuous case). 
	
	{Figures}\ \ref{subfig:persol_prom27a} and \ref{subfig:persol_prom27c} show the corresponding solutions at the points 
	indicated in Fig.~\ref{subfig:persol_m27}. At the beginning, the solution looks like localized states separated by a finite distance. As the norm increases, it gradually delocalizes and forms long stretches of periodic oscillations enclosed by fronts as shown in Fig.\ \ref{subfig:persol_prom27c}. Both cases of localization and delocalization are equivalent to a single localized state in a finite domain. This explains the slanted snaking diagrams observed in Fig.\ \ref{fig:persol_m14m25m26m27} (see also Fig.\ \ref{fig:persol_m14b3} that will be discussed later) \cite{add1}. When the solution becomes completely oscillating, the existence curve stops snaking.

	Comparing the panels in \ref{subfig:persol_m14} and \ref{subfig:persol_m27}, one can note that the complexity of the snaking in the bifurcation curves does not depend on the discretization parameter $h$. To study how the bifurcation curve in one of the panels changes into the other, one would normally vary the parameter $h$. However, in the present numerical setup it may not be possible because we fix the number of sites following the periodicity of the solution, which depends on $h$. To be consistent, if we were to vary the parameter we would also need to change the number of grid points, which can be non trivial to do. In the infinite domain, the change may be related to the attachment or detachment of some parts of the bifurcation curves. 
	
	
	The presence of multiple Maxwell points due to the vanishing of the energy function of the periodic solutions seems to be related to the snaking. To understand the appearance of the additional Maxwell points, it is easier to study them through varying $b_3$ than $h$, which is shown in Fig.\ \ref{fig:persol_m14b3}. The existence curve that initially  only has two Maxwell points are seen to have four Maxwell points in Fig.\ \ref{subfig:persolm14b3_15} as $b_3$ increases. Such an addition  occurs from the tip of a turning point, i.e.,\ a saddle-node bifurcation.
	
	\subsection{Periodic {solutions} for $h\geq2$}
	\begin{figure}[h!]
		\centering
		\includegraphics[scale=0.51]{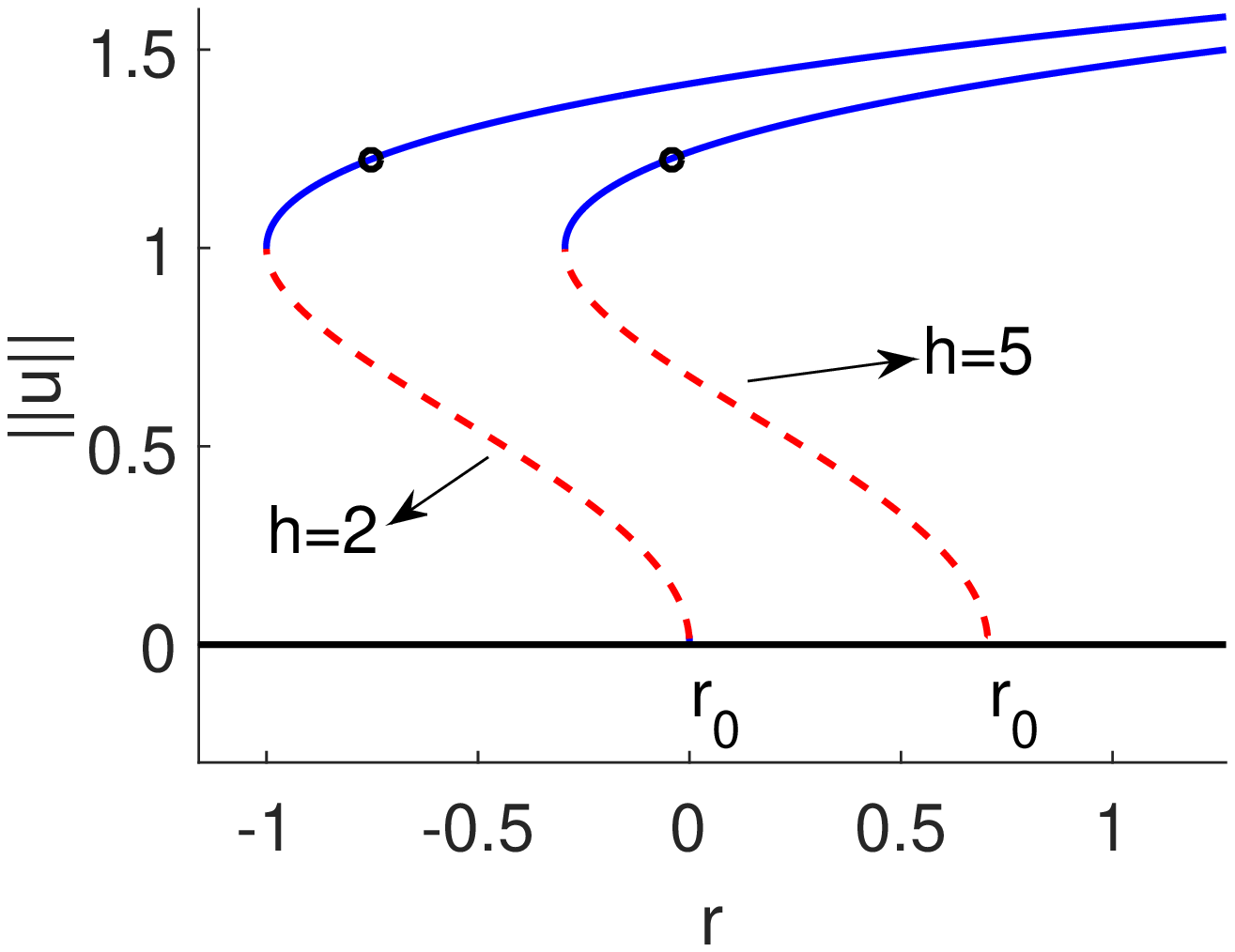}
		\caption{Bifurcation diagram of the periodic solutions for $h\geq2$.
		}
		\label{fig:persol_hgeq2}
	\end{figure}
	
	{Figure}\ \ref{fig:persol_hgeq2}  shows the bifurcation diagram of two periodic solutions for $h\geq2$. Note that in this case, 
	the wave number is always $\pi$ as given in \eqref{eq:khg2} and hence $P=2$.
	The parameter only causes the bifurcation point $r_0$ to shift to the right.
	As $h\rightarrow\infty$, the bifurcation diagram will be equivalent to that of the uniform solution, see Fig.~\ref{fig:unisol_hg2}.

	We can obtain the Maxwell point $r_{M1}$ exactly by equaling the energy of the zero and the periodic solutions \eqref{eq:dSwift-Hohenberg_ti} and \eqref{eq:A_hg2} to yield
	\begin{eqnarray}
	r_{M1}=1-\frac{8}{h^2}+\frac{16}{h^4}-\frac{3b_3^2}{16b_5}.
	\label{eq:Mp_hg2}
	\end{eqnarray}

	\section{Localized solutions}
	\label{sec4}
	In the continuum limit $h\to0$, there are localized solutions bifurcating from $r_0$  \cite{Burke2007a}. In the following, we 
	study the effect of the dicretization to such solutions.

	As derived in Appendix, localized solutions of the discrete Swift-Hohenberg equation bifurcating from $r_0$ are given asymptotically at the leading order by	
	\begin{eqnarray}
	{u}_{l,n}&=&\sqrt{\frac{2(r_0-r)}{3b_3}}\sech\left(hn\left(\frac{(r-r_0)}{C} \right)^{\frac{1}{2}}\right)\nonumber\\
	&&\times\cos\left(khn+\phi\right)+\mathcal{O}(r-r_0).\label{eq:Un_loc}
	\end{eqnarray}
	
	Note that the parameter $\phi$ is the phase of the
	pattern within the sech envelope, which within this asymptotics remains arbitrary. In the continuum limit $h\to0$, the phase-shift is $\phi=0$ or ${\pi}/{2}$ \cite{Burke2007a}, which can only be determined using exponential asymptotics \cite{Chapman2009,Kozyreff2006,Susanto2011,Matthews2011}. Here, aside from the locking between the sech envelope and the underlying wave train, for $h>0$ there is also the possibility for the envelope to be locked with the spatial discretization. However, this will be beyond the scope of this paper and in the following we will only consider the phase pertaining to the continuous limit above. 
	
	\begin{figure}[htbp!]
		\centering
		\subfigure[]{\includegraphics[scale=0.51]{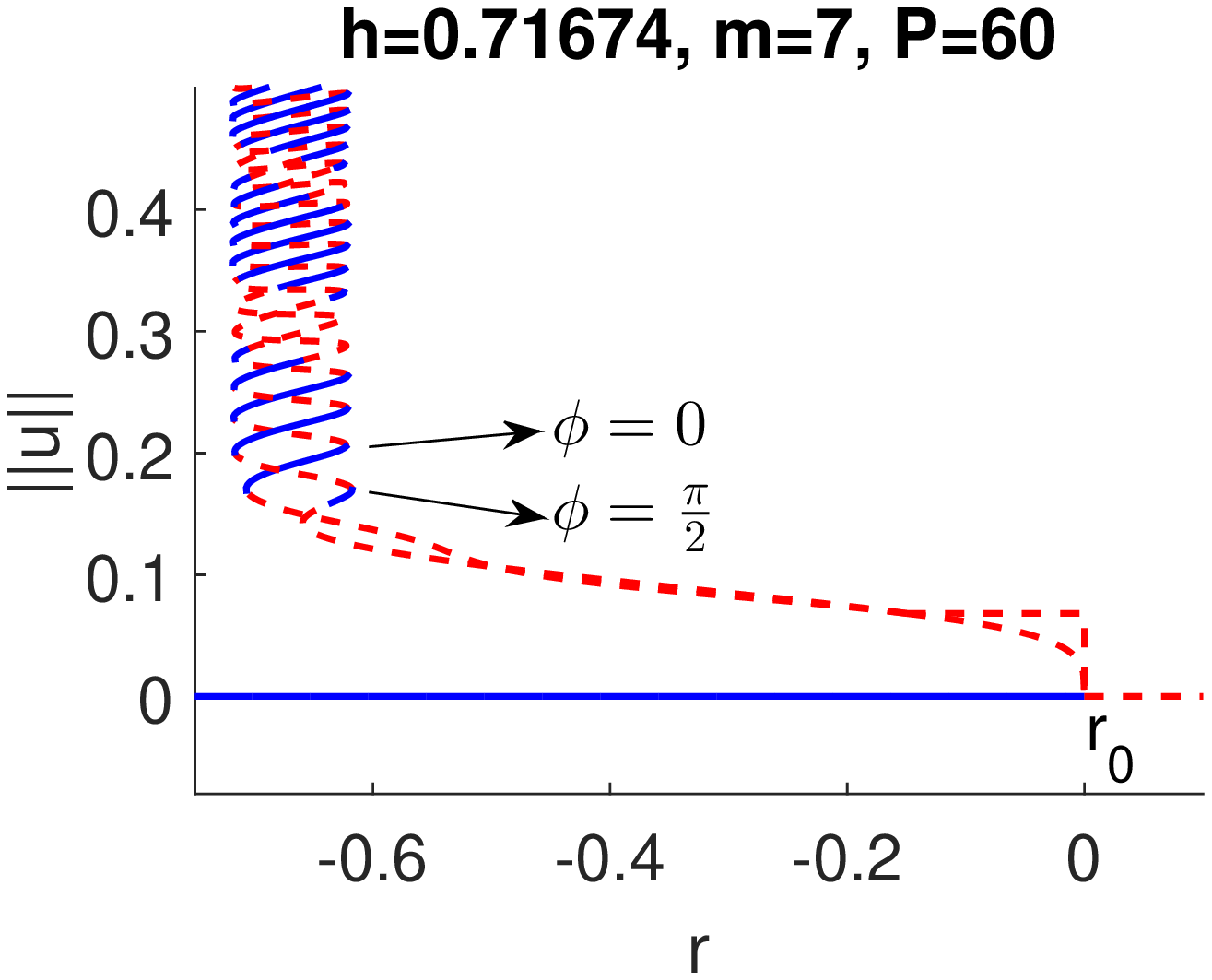}\label{subfig:Ls_m7}}
		\subfigure[]{\includegraphics[scale=0.285]{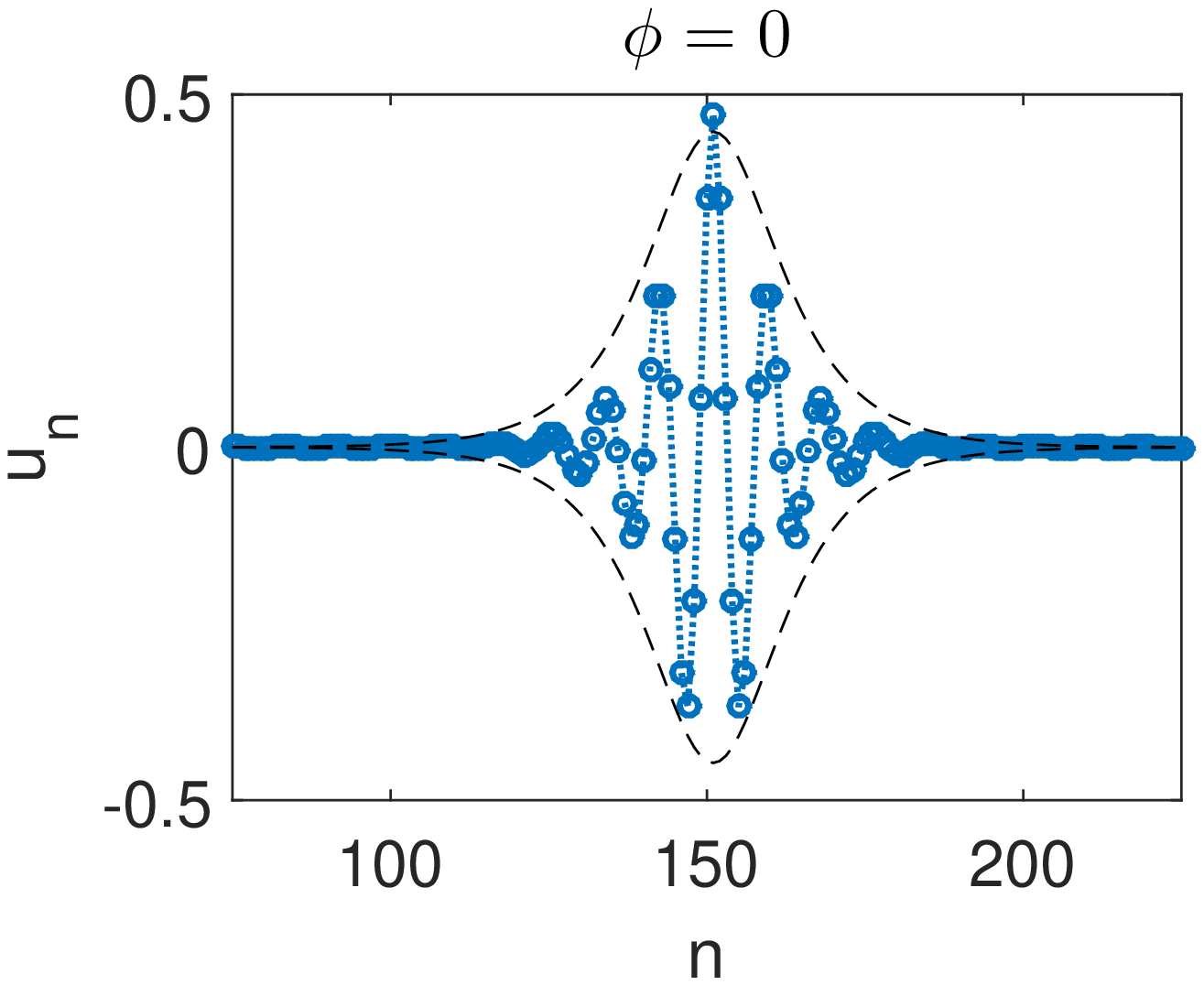}\label{subfig:Lsp_m71a}}
		\subfigure[]{\includegraphics[scale=0.285]{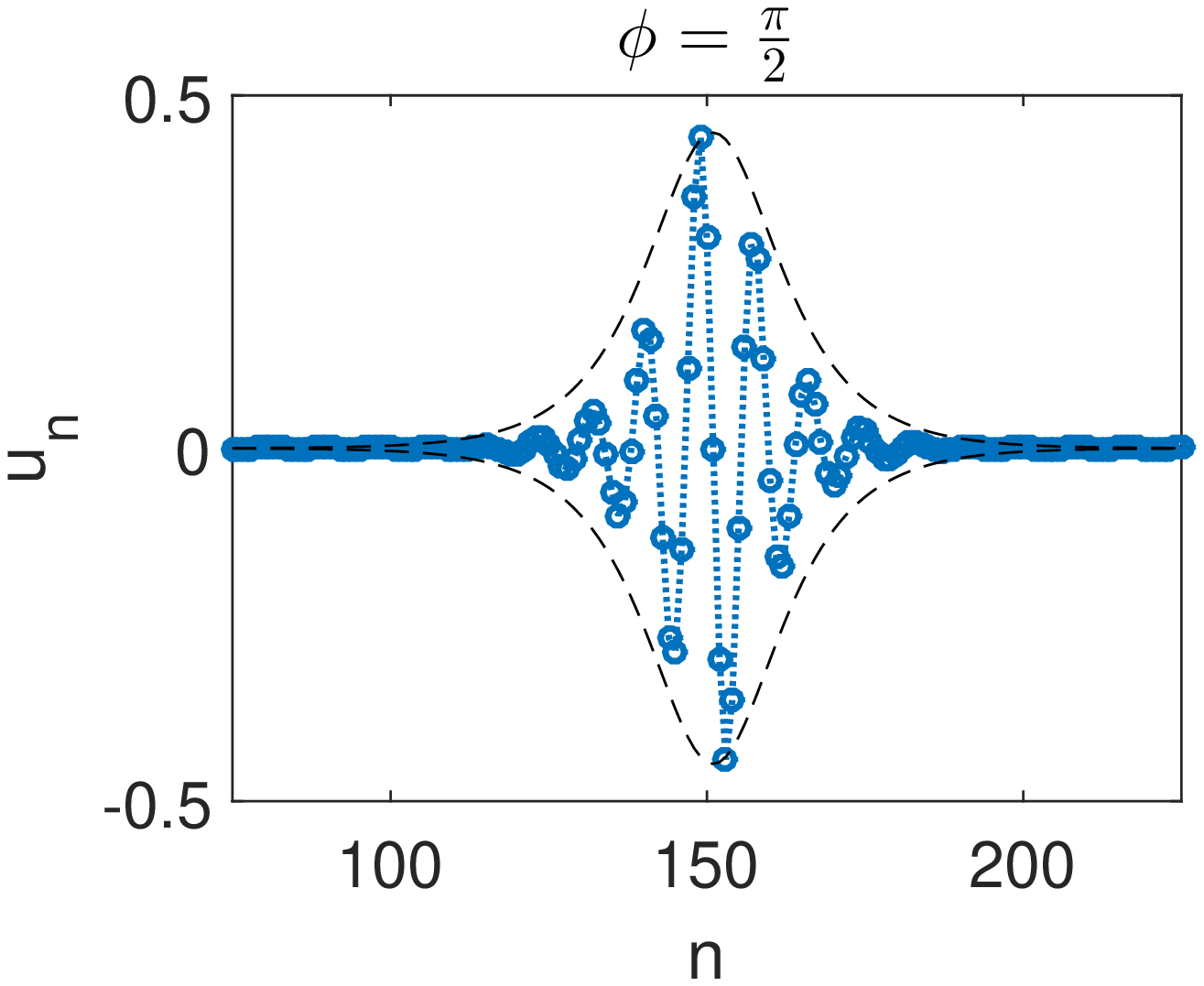}\label{subfig:Lsp_m72b}}
		\caption{(a) Bifurcation diagram of the localized solution for $h=0.71674$ with $m=7,\, P=60.$ {(b), (c)} Profiles of localized solutions next to the bifurcation point $r_0$. The dashed curves in panel {(b), (c)} {correspond to} the envelope given by \eqref{eq:Un_loc}. 
				}
		\label{fig:Ls_m5m7}
	\end{figure}	
	
	\subsection{Snaking regions: $r$ vs.\ $h$}
	\label{sec:hl1} 

	By using Eq.\ \eqref{eq:Un_loc} as our initial guess for the numerics, we obtain the existence curve of localized solutions. {Figure}\ \ref{fig:Ls_m5m7} shows the bifurcation diagram of the localized solutions that form the snaking behavior for the phase-shift $\phi=0$ and ${\pi}/{2}$ for a value of $h<1$. 
	The vertical axis is the solution norm [see \eqref{eq:norm_d}].
	We also show the corresponding solutions in the same figures.
		
	One can note that the bifurcation diagram is {similar} to that of the continuous Swift-Hohenberg equation \cite{Burke2007a}. However, we note one difference where up in the snaking diagram, we obtain intervals of norm where both solutions are all unstable.

	\begin{figure}[htbp!]
		\centering
		\includegraphics[scale=0.56]{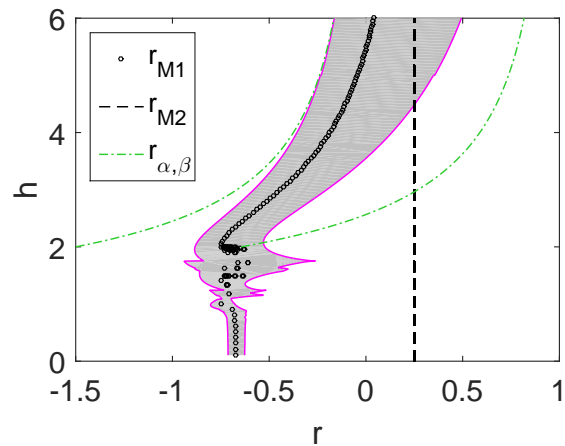}
		\caption{{The pinning region as a function of $h$ indicated by the gray area. The Maxwell points are also denoted. $r_{M2}$ is defined in \eqref{eq:Mp_2nd}. $r_\alpha$ and $r_\beta$ are analytical approximations derived in Sec.\ \mbox{\ref{aap}}.}
		}
		\label{fig:prb3_2}
	\end{figure}

	
	At present we may conclude that the discretization parameter $h$ only {effects slightly} to the snaking behavior. However, in the following we will show that the range of parameter $1\leq h<2$ is particularly peculiar as there are detachments of snaking structures.

\begin{figure}[t!]
		\centering
		\subfigure[]{\includegraphics[scale=0.55]{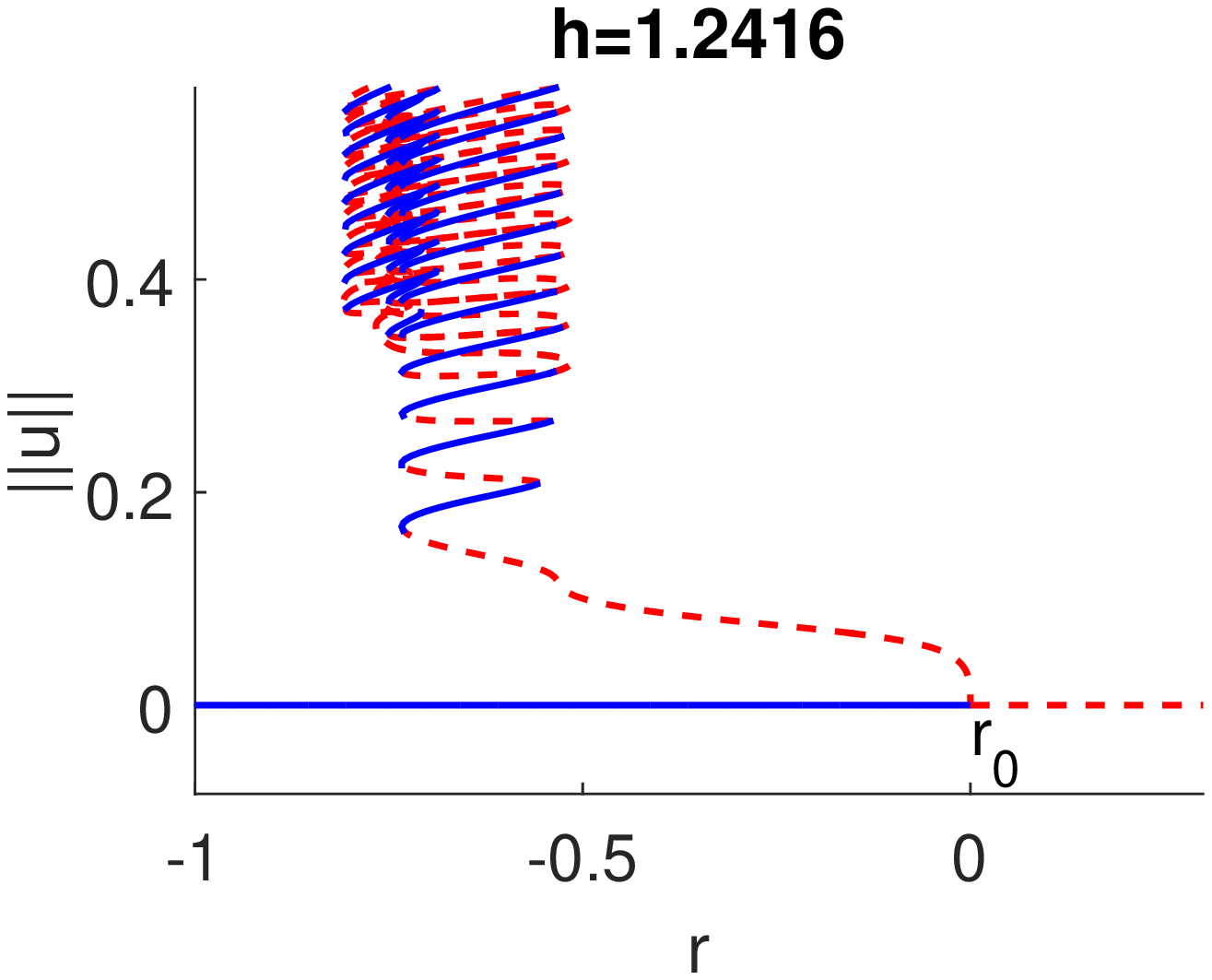}\label{subfig:jump_br1}}
		\subfigure[]{\includegraphics[scale=0.55]{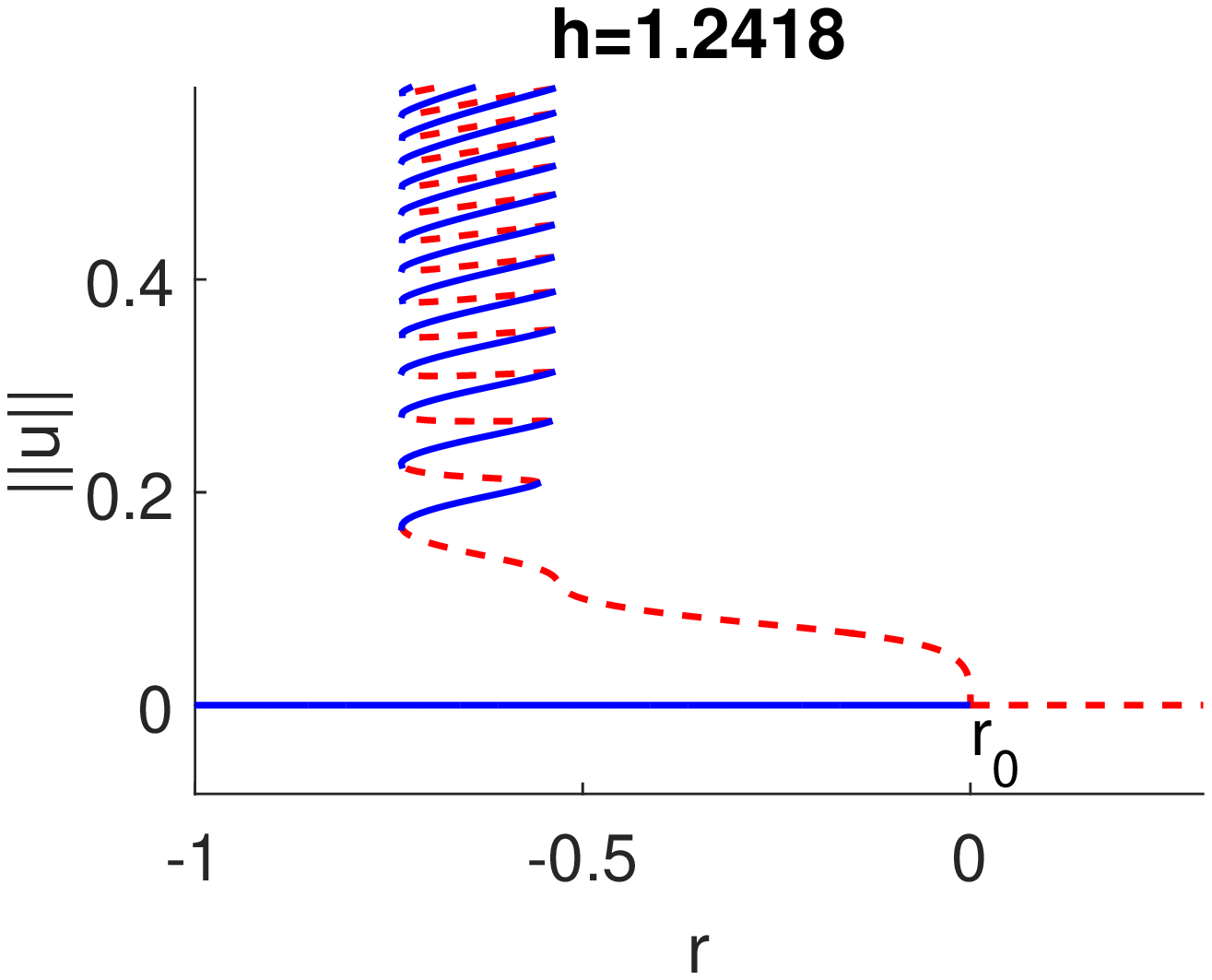}\label{subfig:jump_br2_2}}
		\caption{The homoclinic snaking {(a)} before and {(b)} after a jump in Fig.\ \ref{fig:prb3_2}.}
		\label{fig:jump_br}
	\end{figure}

	In Fig.~\ref{fig:prb3_2} we plot the pinning region, {which is bounded by left and right turning points of the snaking curve}, for varying $h$. We obtain smooth boundaries in the regions $h<1$ and $h\geq2$, while there are jumps and pikes of pinning region boundaries in the region $1\leq h<2$. Analyzing the snaking profiles around the jump or spiking points closely, we obtain that they  correspond to the detachment of a snaking profile from the main branch  as depicted in Fig.~\ref{fig:jump_br}.
	
	\begin{figure}[tbp!]
		\centering
		\includegraphics[scale=0.56]{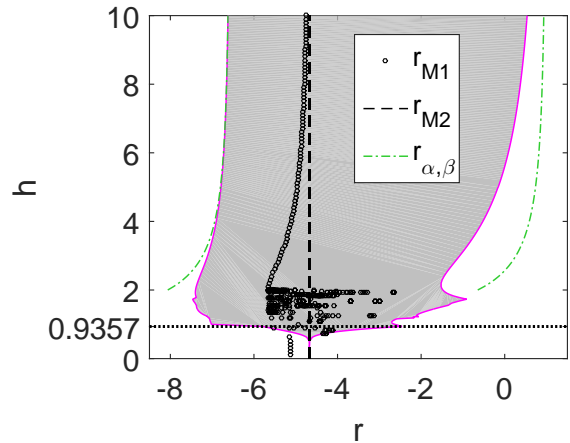}
		\caption{The same as in Fig.\ \ref{fig:prb3_2} for $b_3=5.5$. The horizontal dotted line indicates a sample value of $h$ that will be considered further later (see Fig.\ \ref{fig:prof_b3_55}) for having three types of localized solutions.}
		\label{fig:prb3_55}
	\end{figure}

	\begin{figure*}[htbp!]
		\centering
		\subfigure[]{\includegraphics[scale=0.525]{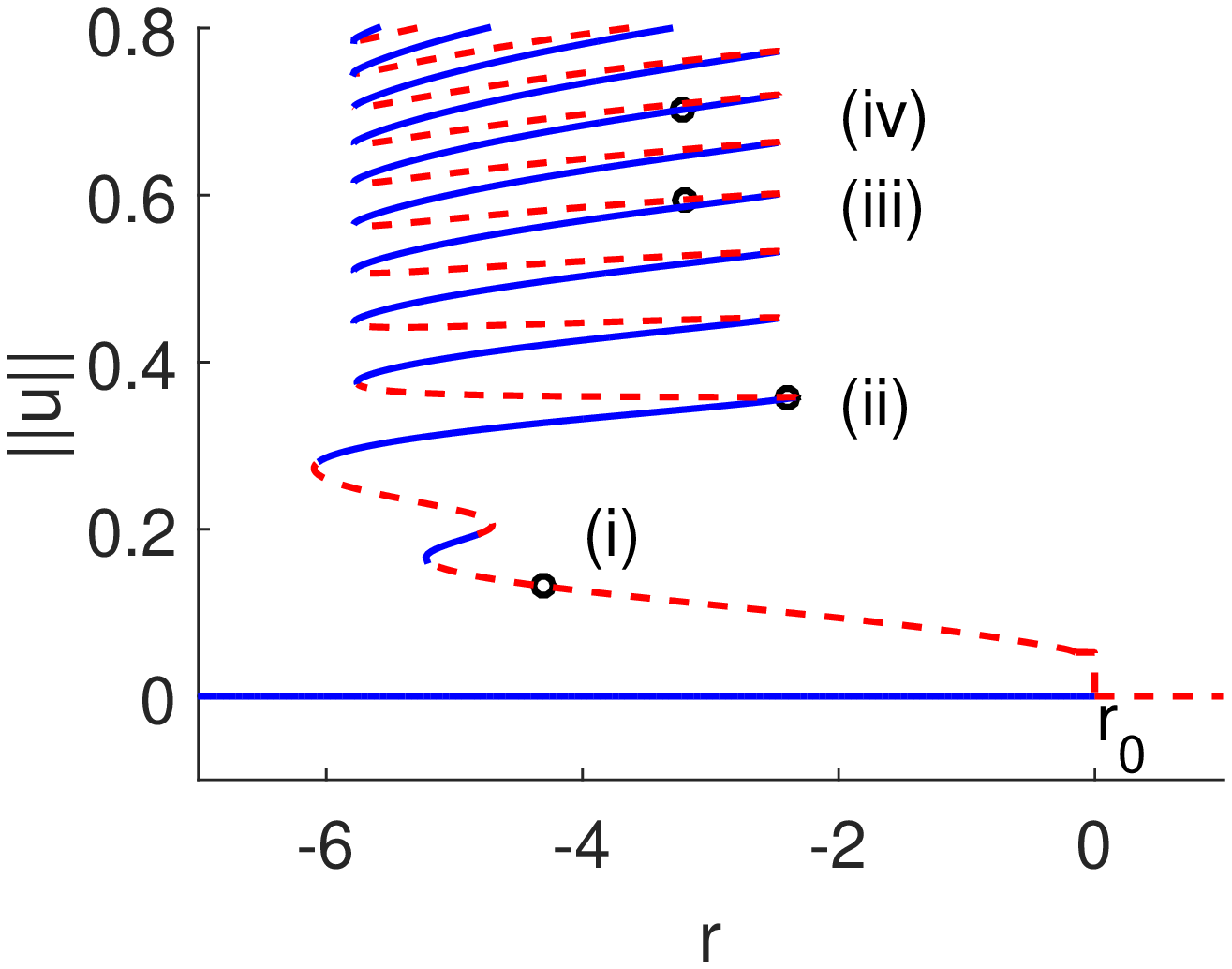}\label{subfig:loc_sol21}}
		\subfigure[]{\includegraphics[scale=0.525]{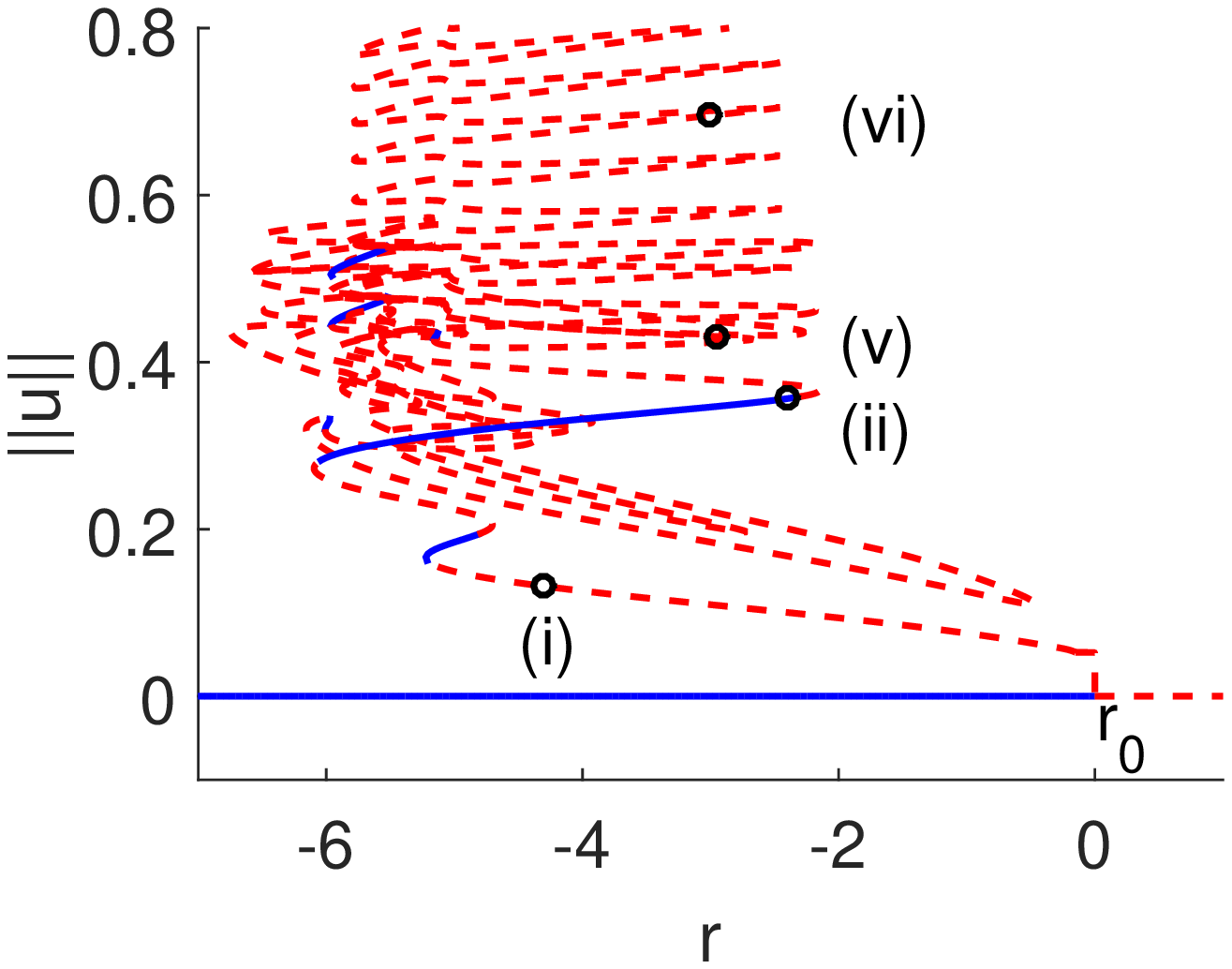}\label{subfig:loc_sol2}}
		\subfigure[]{\includegraphics[scale=0.525]{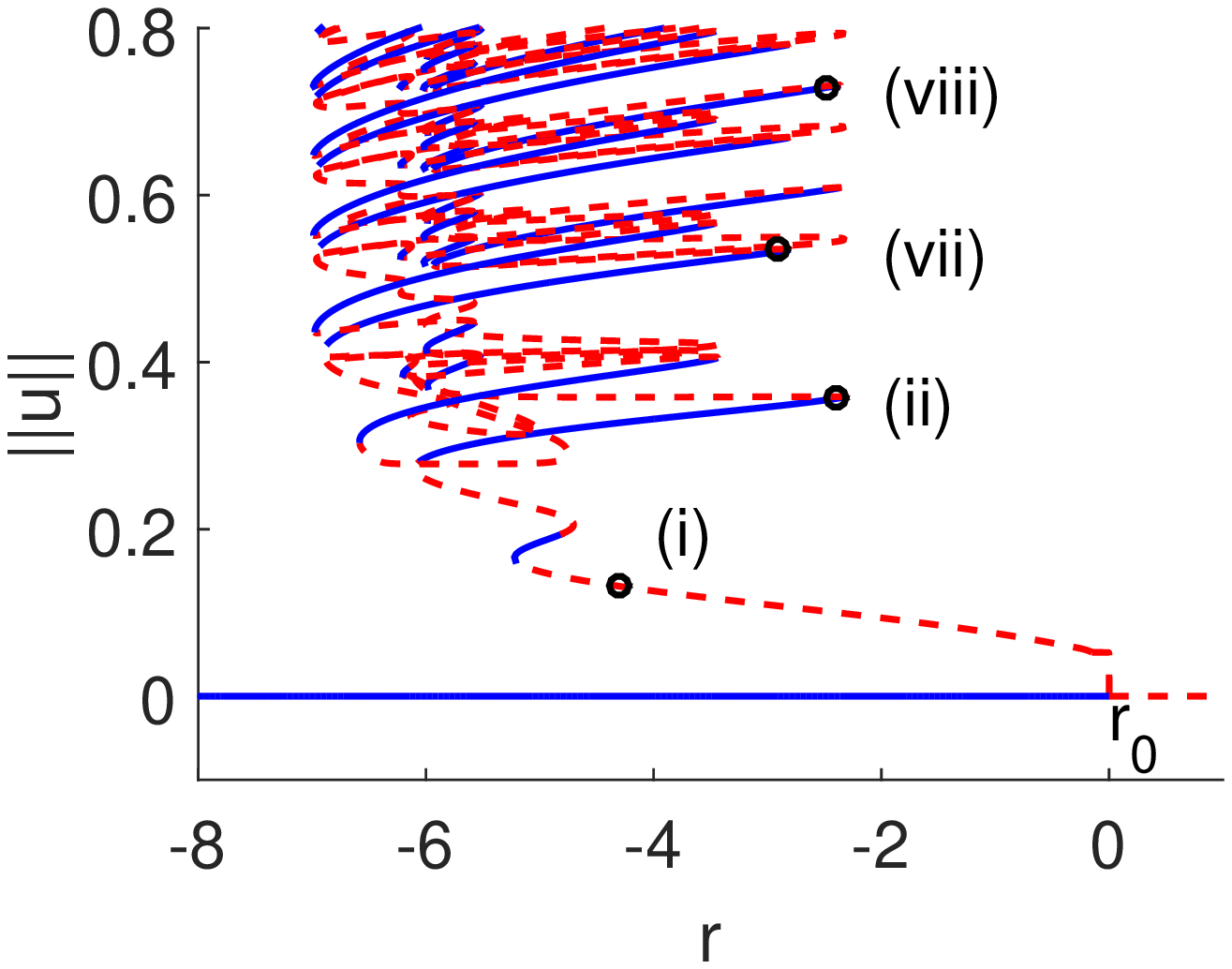}\label{subfig:loc_sol3}}
	\end{figure*}
	\begin{figure*}[htbp!]
		\centering
		\subfigure[]{\includegraphics[scale=0.5]{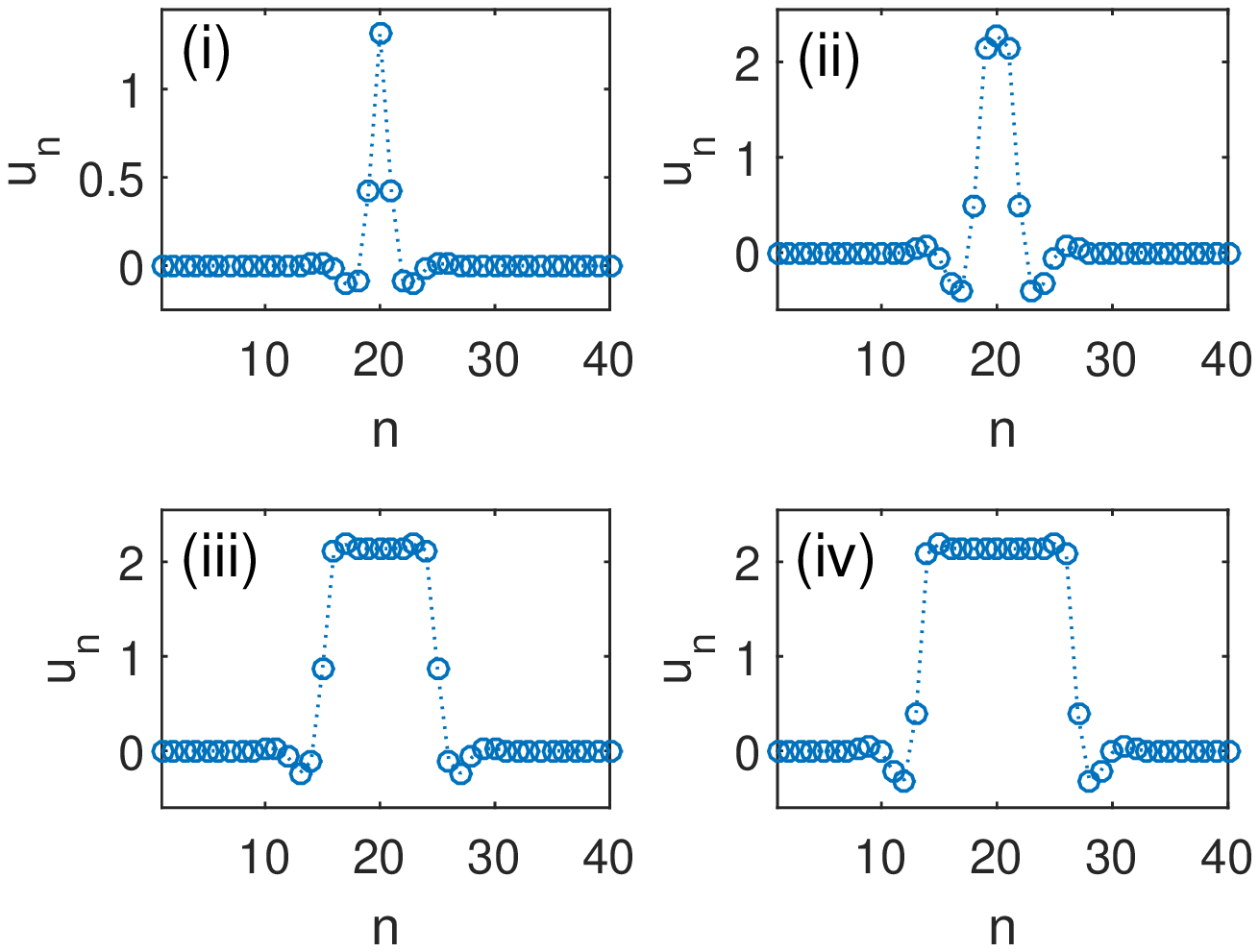}\label{subfig:prof_b3_55_1}}		
		\subfigure[]{\includegraphics[scale=0.5]{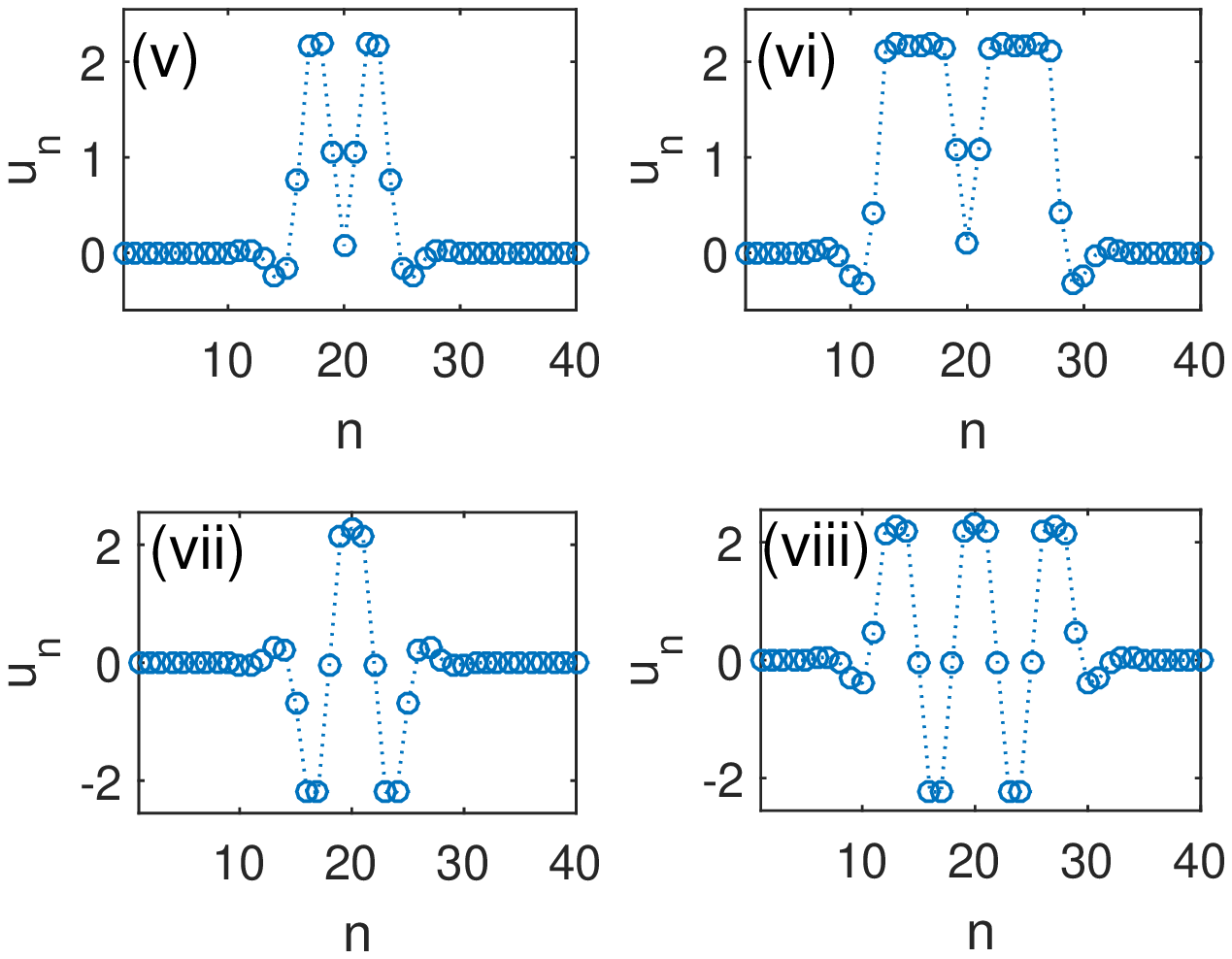}\label{subfig:prof_b3_55_2}}		
		\caption{{Three different snaking diagrams (a)-(c) and their localized solutions (d) and (e) that were obtained for $h=0.9357$ with $\phi=0$ and $b_3=5.5$. The three diagrams share the same portion of curve between $r_0$ and point (ii). In the continuum limit $h\to0$, only the bifurcation diagrams in (a) and (b) were reported in \mbox{\cite{Burke2006}}.			}}
			\label{fig:prof_b3_55}
	\end{figure*}

	\begin{figure}[tbp!]
		\centering
		\subfigure[]{\includegraphics[scale=0.51]{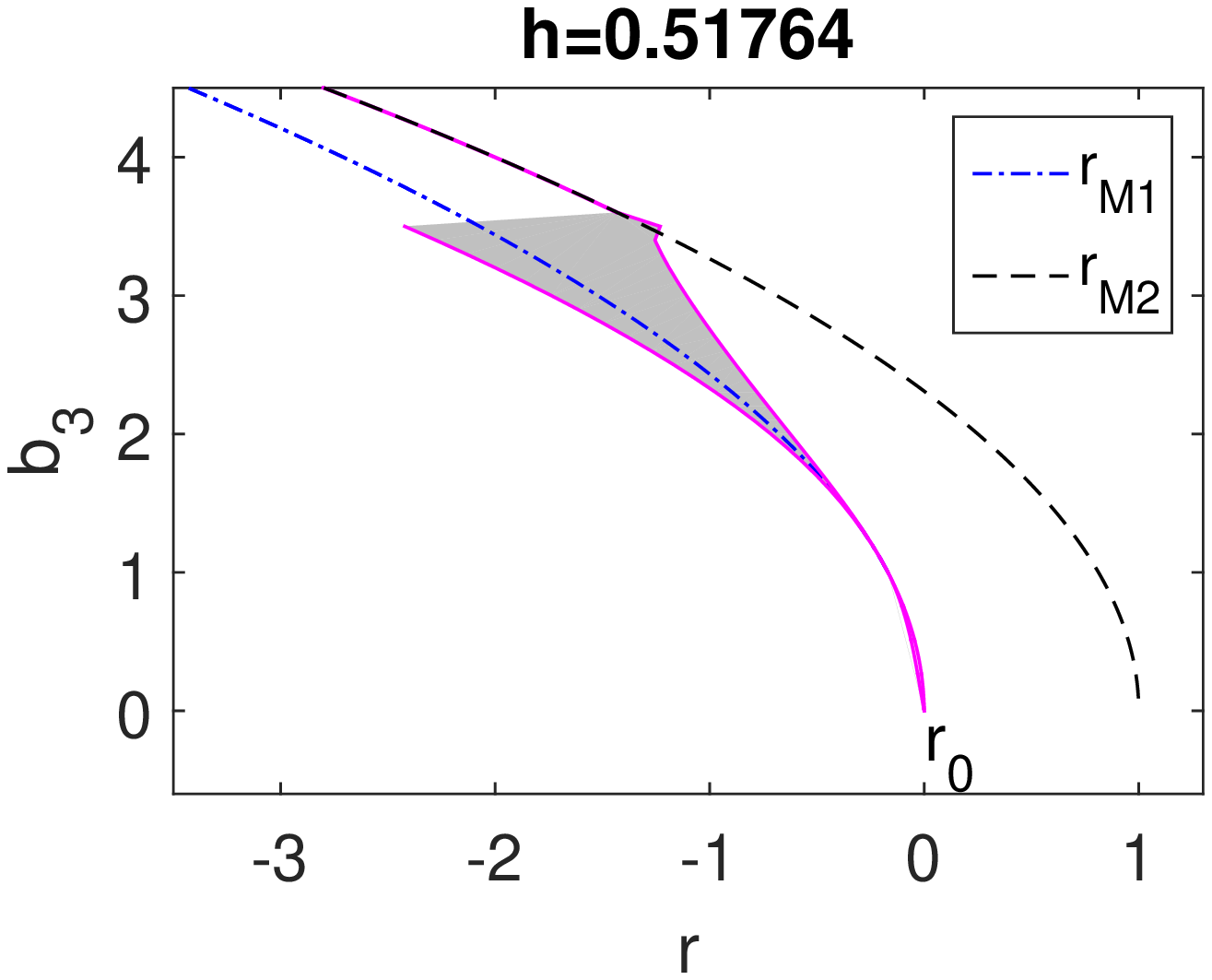}\label{fig:prhl1}}
		\subfigure[]{\includegraphics[scale=0.51]{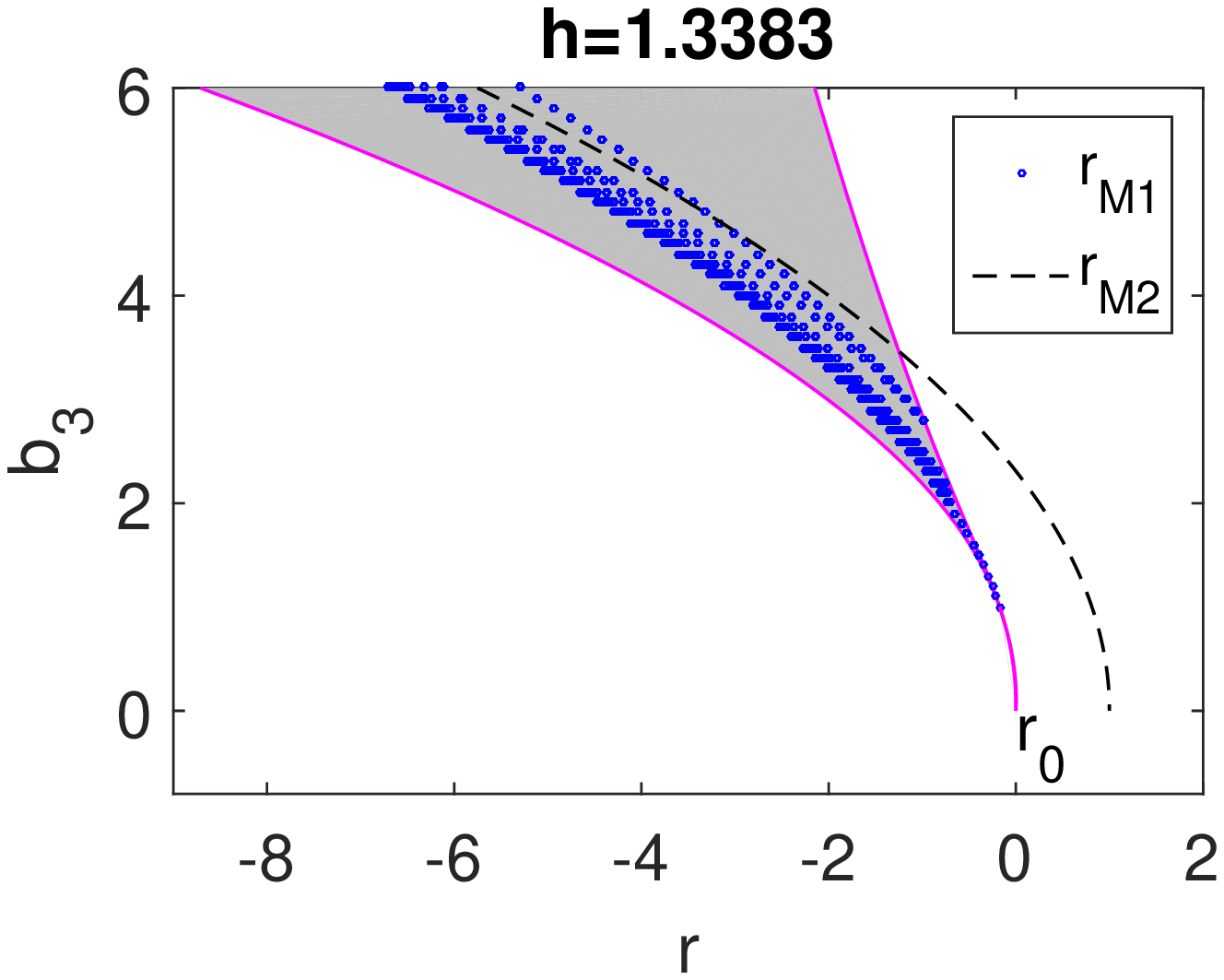}\label{fig:prhl2}}
		\subfigure[]{\includegraphics[scale=0.51]{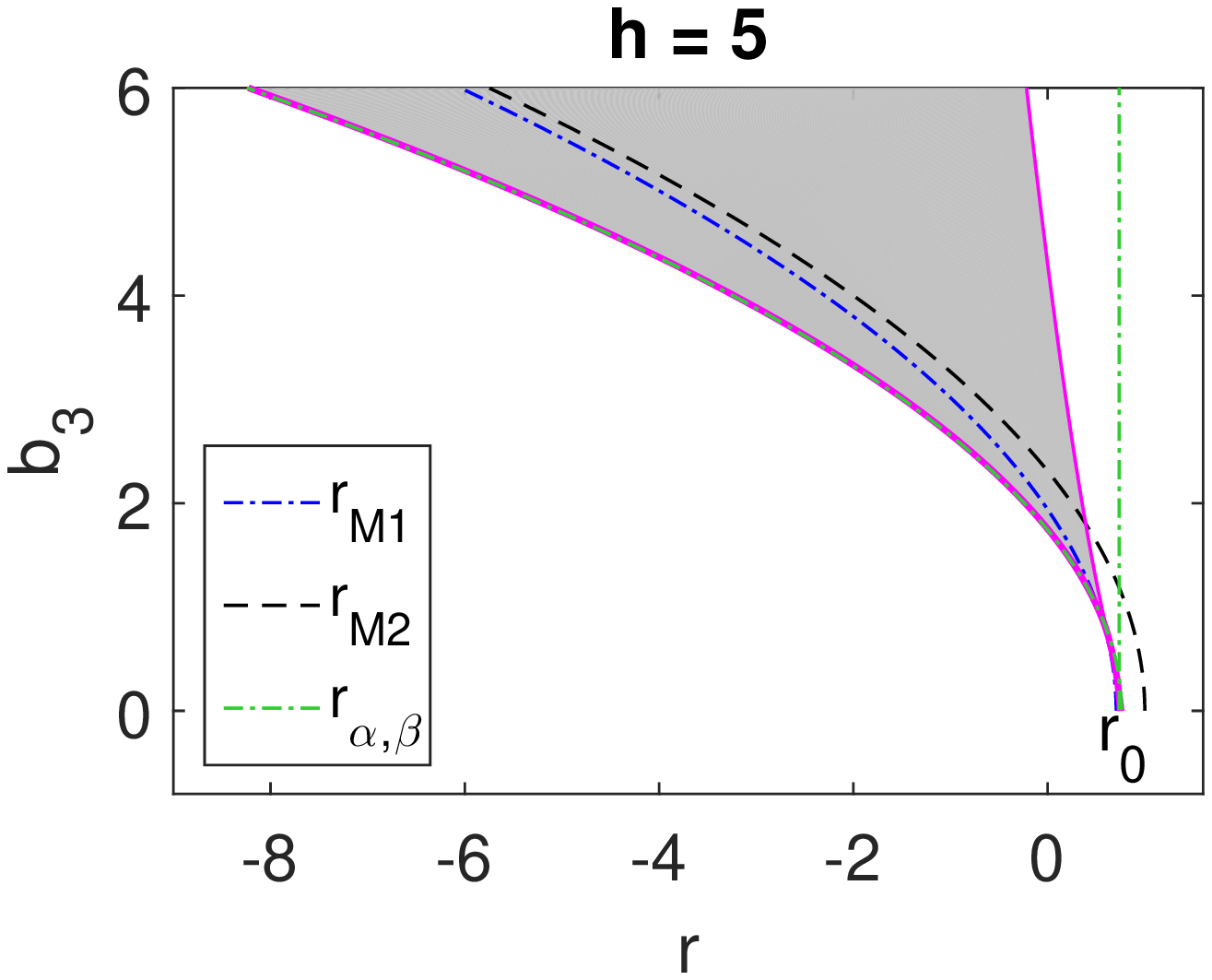}\label{fig:prhg2}}
		\caption{The pinning regions indicated by the gray area for three different values of $h$ representing the region (a) $h\leq1$, (b) $1\leq h<2$, and (c) $h\geq 2$. See the text for the definition of $r_{M1}$, $r_{M2}$, $r_{\alpha}$, and $r_{\beta}.$}	
	\end{figure}
	
	{Figure}\ {\ref{subfig:jump_br1}} shows the bifurcation diagram before the jump, showing a complex snaking. 
	Right after the jump, we obtain a much simpler snaking structure as shown in Fig.\ \ref{subfig:jump_br2_2}.
	The change of the bifurcation diagram in panel (a) to that in panel (b) is due to the detachment of a snaking structure (not shown here) from the main branch, which is the general scenario of the jumps and pikes observed in Fig.\ {\ref{fig:prb3_2}}.

	Overall we say from Fig.\ \ref{fig:prb3_2} that	for $h<1$, the influence of the discretization from the fourth derivative term is more dominant than the second derivative one, while for $h\geq2$ it is the opposite. For the intermediate interval $1\leq h<2$, the influence of the discretization from the second and the fourth derivatives is relatively the same that yields non-trivial bifurcation curves.
	
	In Fig.\ \ref{fig:prb3_2}, we depict Maxwell points defined previously as the points when the energy of the periodic solutions vanishes. It is particularly interesting to note that there are many Maxwell points in the region $1\leq h <2$, especially when $h\to2$. Additionally, we also plot $r_{M2}$ as vertical dashed line, that is defined as the point when the energies of {$U_+$} and {$U_0$} are equal, i.e.,\ $E[U_+]-E[U_0]=0$. The point can be calculated easily as
	\begin{eqnarray}
	r_{M2}=1-\frac{3b_3^2}{16b_5},\label{eq:Mp_2nd}
	\end{eqnarray}
	which is exactly the same as that of the continuum limit \cite{Burke2007a}. This special point will also be relevant later on when we consider the effect of varying $b_3$.

	In the continuous case, it was shown that the pinning region enlarges with increasing $b_3$ and above a critical $b_3\approx3.521$ there is no snaking formed any more \cite{Burke2007a}. The snaking simply just collapses into a vertical line. This happens when the right boundary of the snaking region touches the special point \eqref{eq:Mp_2nd}. In Fig.\ \ref{fig:prb3_55} we plot the pinning region for varying discretization parameter $h$ with $b_3=5.5$. One can observe that for small $h$ indeed there is no snaking. However, when $h$ is large enough, a snaking behavior is obtained again.
	
	For this value of $b_3$, we also still see jumps and pikes along the pinning region boundaries. In this case we even observe a more complicated structure than that in Fig.\ \ref{fig:jump_br}, where the snaking involves three different branches. One example is for $h\sim0.9357$. We show in Fig.\ \ref{fig:prof_b3_55} the different branches and their corresponding solutions.

	{The three bifurcation diagrams in Figs.\ \ref{subfig:loc_sol21}, \ref{subfig:loc_sol2}, and \ref{subfig:loc_sol3} share the same portion of curves from the bifurcation point $r_0$ until point (ii). Point (ii) is a bifurcation point, from which emanates the three different branches. In Figs.\ \ref{subfig:prof_b3_55_1} and \ref{subfig:prof_b3_55_2} we show the corresponding solution profiles for each branch at the indicated points in Figs.\ \ref{subfig:loc_sol21}, \ref{subfig:loc_sol2}, and \ref{subfig:loc_sol3}.}

	For $h<0.9357$, the solutions are similar to those in Fig. \mbox{\ref{subfig:prof_b3_55_1}}, i.e.,\ branch in Fig.\ \ref{subfig:loc_sol21}. For $h\approx0.9357$, the solutions in Figs.~\mbox{\ref{subfig:prof_b3_55_1}} and~\mbox{\ref{subfig:prof_b3_55_2}} coexist. In particular, localized solutions 
	such as those shown in Fig.~\mbox{\ref{subfig:prof_b3_55_2}} are the ones that give a complicated bifurcation diagram that experiences detachment and attachment processes for $0.9357\leq h <2 $.
	
	{Solutions in Figs.~\mbox{\ref{subfig:prof_b3_55_1}} and\
		\mbox{\ref{subfig:prof_b3_55_2}} can be seen to rather have a flat plateau around $U_+$, from which one obtains their relation to the special point $r_{M2}$ \mbox{\cite{Burke2006}}. In the continuum limit $h\to0$, the reported diagram was only that shown in Figs.~\ref{subfig:loc_sol21} and \ref{subfig:loc_sol2} \cite{Burke2006}. 
		 
}
	

	\subsection{Pinning regions: $\displaystyle r$ vs.\ $b_3$}
	\label{sec5}
	
	Here, we would like to study further the effect of the parameter $b_3$ on the snaking in the discrete system. We now fix $h$ and vary $b_3$ instead.	
	
	{Figure}\ \ref{fig:prhl1} shows the pinning region for $h<1$ represented by $h=0.51764$. The region behaves quite similarly as the continuous Swift-Hohenberg equation \cite{Burke2006}. Maxwell point is always inside the snaking region. Beyond $b_3=3.521$, the solution stops snaking and follows the  point $r_{M2}$ \cite{Burke2007a}.

	{Figure}\ \ref{fig:prhl2} shows the pinning region for $1\leq h<2$, which is represented by $h=1.3383$. The discretization causes the presence of multiple Maxwell points 
	appearing inside the pinning region. What is notable is the result that unlike the previous case for $h<1$, here the pinning region does not feel the presence of the special point $r_{M2}$.

	{Figure}\ \ref{fig:prhg2} shows the pinning region for $h\geq2$, which is represented by $h=5$. The result shows that there is only one Maxwell point (\ref{eq:Mp_hg2}). 
	Note that $r_{M1}$ and $r_{M2}$ converge to the same point when $h\rightarrow \infty$. The point $r_{M2}$ does not affect the pinning region just like in the previous case when $1\leq h<2$. 

	\section{Analytical approximation}\label{aap}
	
\begin{figure}[h!]
		\centering
		\subfigure[]{\includegraphics[scale=0.55]{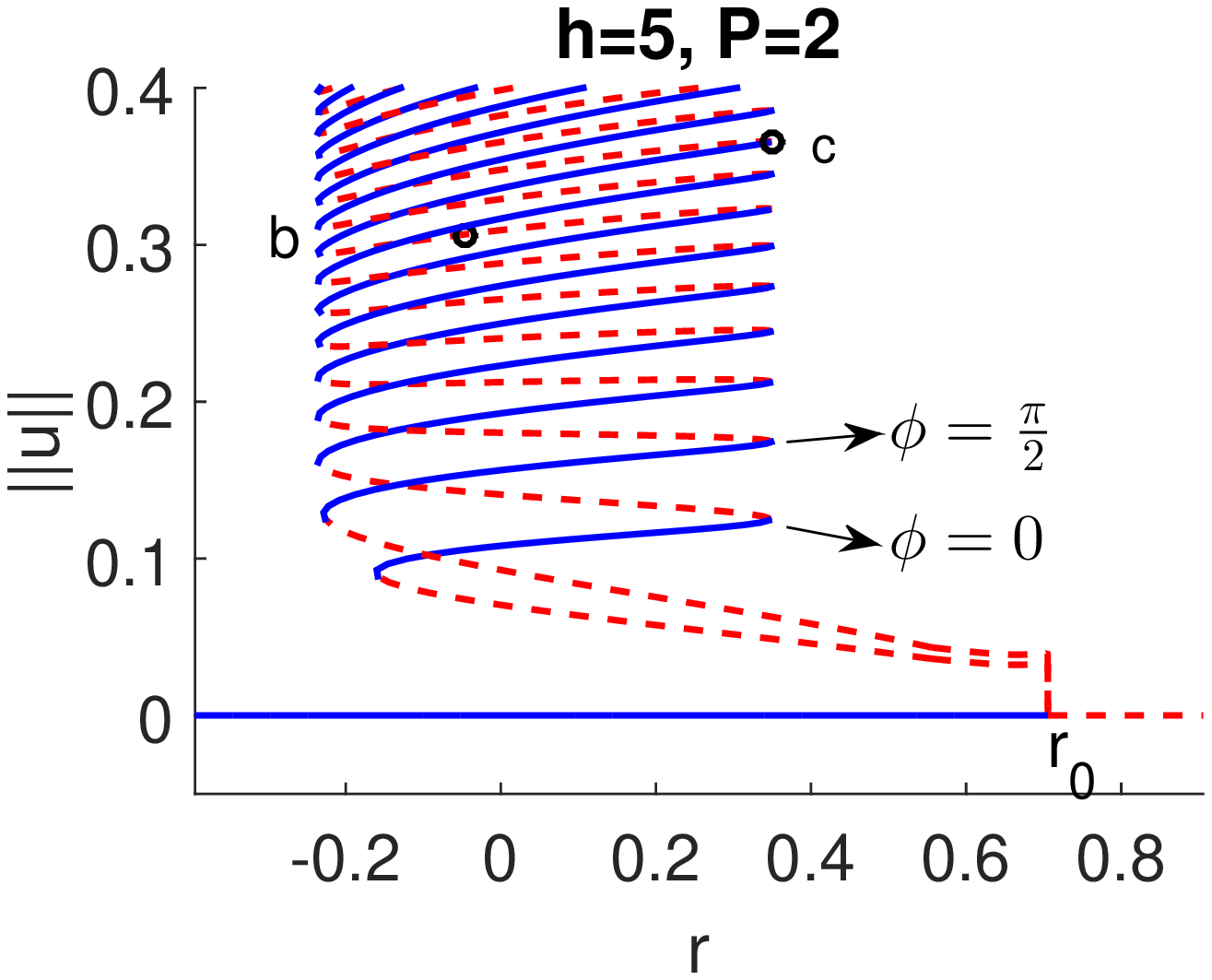}\label{subfig:Ls_h5}}
		\subfigure[]{\includegraphics[scale=0.286]{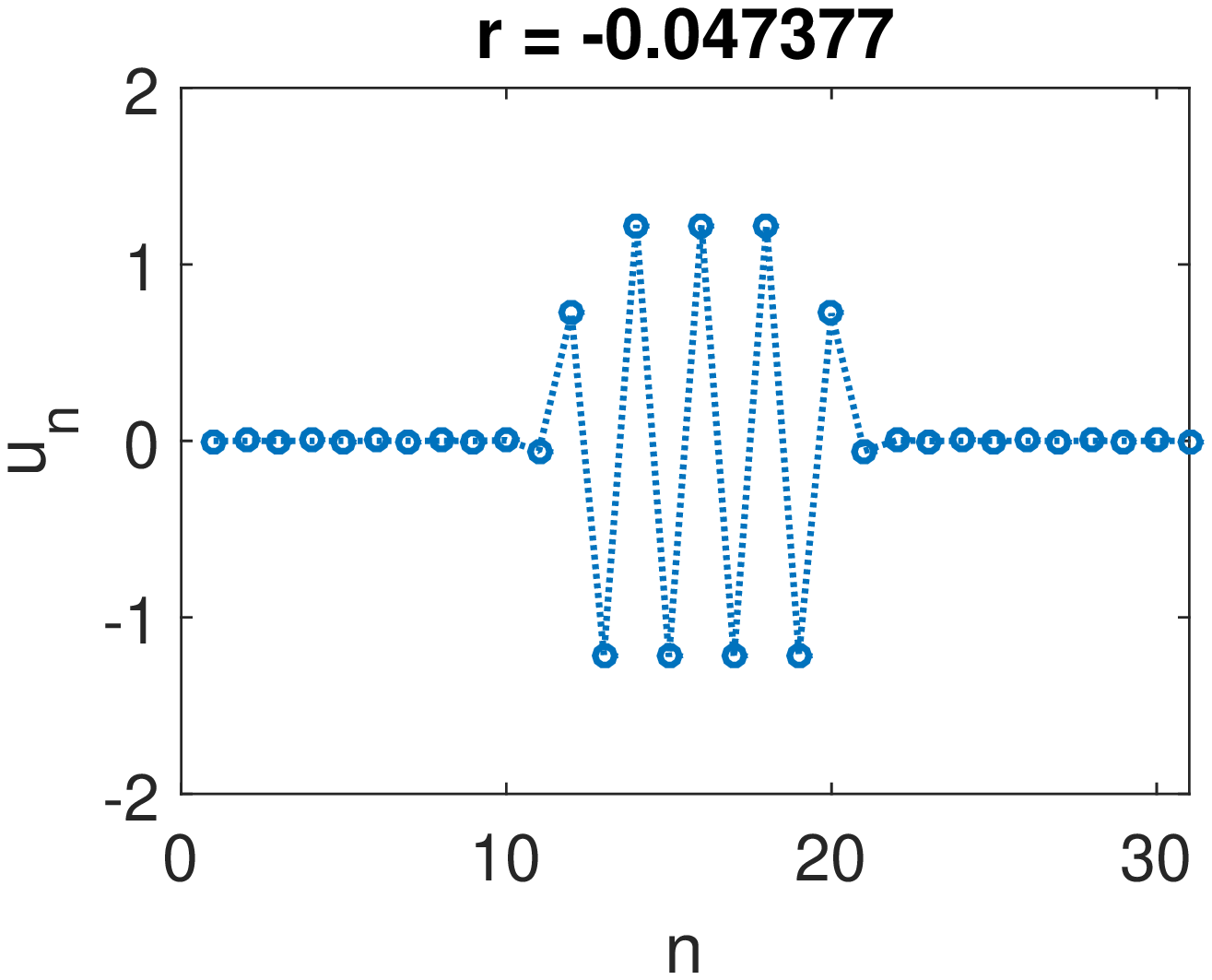}\label{subfig:prof_h5_b}}
		\subfigure[]{\includegraphics[scale=0.286]{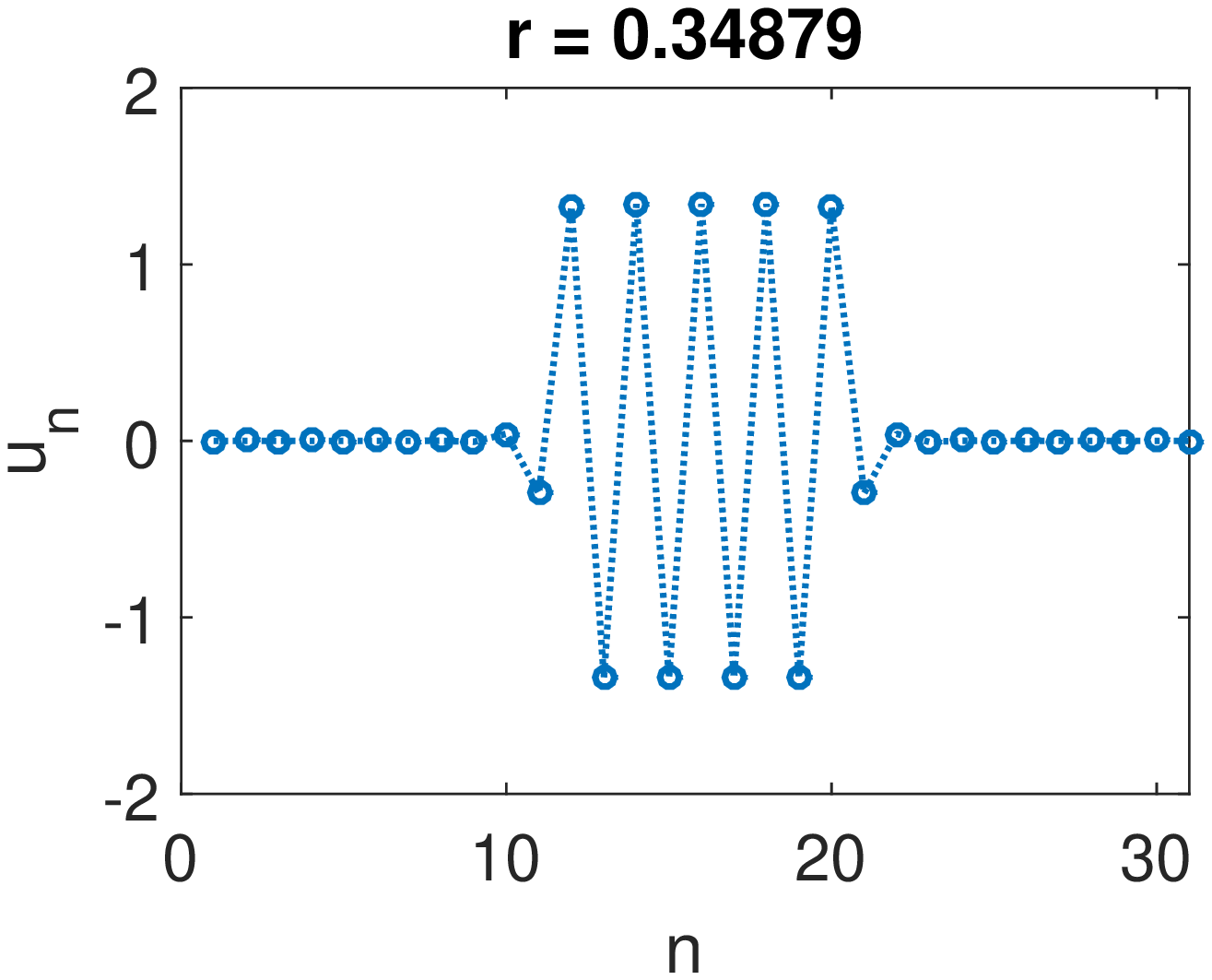}\label{subfig:prof_h5_d}}	
		\caption{{The bifurcation diagram of localized solutions for {$h=5$} and their corresponding profiles at the points indicated by the letters in (a).}}
		\label{fig:prof_pos_h5}
	\end{figure}

	It is important to note that when $h\gg1$, the discrete system is actually weakly coupled. {Figure}\ \ref{fig:prof_pos_h5} shows the bifurcation diagram of localized solutions for $h=5$ and the corresponding solutions. The panels show the fact that as we vary $r$ along the branch, there is basically only one node that is active and varies following the variation of the parameter $r$, while the other points are either in the periodic solution part or in the region of the uniform zero solution.
	
	{From (\ref{eq:dSwift-Hohenberg})}, we can assume that up in the snaking diagram only five nodes are involved in the dynamics, i.e.,\
	\begin{eqnarray}
	&u_{n-2}=0,\quad u_{n-1}=0,\quad  u_n=\upsilon,\quad u_{n+1}=\pm\hat{\varepsilon},& \nonumber\\
	&\text{and} \quad u_{n+2}=\mp\hat{\varepsilon}.&\label{eq:coupling_sh_assume}
	\end{eqnarray}
	Here, $\hat{\varepsilon}$ is the approximate amplitude of the periodic solution given by \eqref{eq:A_hg2} and $\upsilon$ is the active node. 
	Substituting \eqref{eq:coupling_sh_assume} into the time-independent discrete Swift-Hohenberg equation \eqref{eq:dSwift-Hohenberg} will yield a fifth order polynomial for the variable $\upsilon$
	\begin{eqnarray}
	\begin{aligned}
	P_5\left(\upsilon\right)=&-b_5 \upsilon^5+b_3\upsilon^3+\left(r-1+\frac{4}{h^2}-\frac{6}{h^4}\right)\upsilon\\
	&\mp\frac{2\hat{\varepsilon}}{h^2}\pm\frac{5\hat{\varepsilon}}{h^4}=0.
	\end{aligned}
	\label{eq:poly_snake}
	\end{eqnarray}
	\begin{figure}[tbp!]
		\centering
		\includegraphics[scale=0.56]{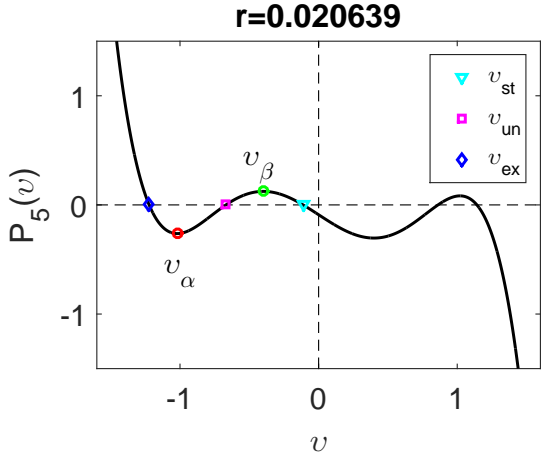}
		\caption{One-active site polynomial for $h=5$.
			$\upsilon_\alpha$ and $\upsilon_\beta$ represents the left and the right pinning boundary.
			$\upsilon_{st}$ represents the stable site of the lower solution.
			$\upsilon_{un}$ represents the unstable site of the solution.
			$\upsilon_{ex}$ represents the stable site of the upper solution.
		}\label{fig:poly_h5}
	\end{figure}

	We call \eqref{eq:poly_snake} a one-active site approximation. Without loss of generality, we can consider one sign only from the plus-minuses in the polynomial because of its symmetry. We plot in Fig.\ \ref{fig:poly_h5} the polynomial \eqref{eq:poly_snake} {for} $h=5$.
	
	In general, the function will have five real roots. Three of them are related	to the snaking as they can disappear in a saddle-node bifurcation with varying $r$. The roots are indicated in Fig.\ \ref{fig:poly_h5}. The boundaries of the pinning region can then immediately be recognized as the condition when the local minimum  at $\upsilon=\upsilon_\alpha$ or the local maximum at $\upsilon=\upsilon_\beta$ touches the horizontal axis. To be precise, $\upsilon_\alpha$ and $\upsilon_\beta$ correspond to the left and right {boundaries} of the pinning region, respectively.
	
	It is rather straightforward to obtain that
	\begin{flalign}
		\upsilon_{\alpha,\beta}=&\,\frac{1}{10b_5h}\left(10b_5\left(3b_3h^2\pm\left(h^4\left(20b_5\left(r-r_0\right)+9b_3^2\right)\right.\right.\right.\nonumber\\ &\left.\left.\left.-80h^2b_5+200b_5\right)^\frac{1}{2}\right)\right)^{\frac{1}{2}}\nonumber\\
		=&\,\frac{1}{\sqrt{10b_5}}\left(3b_3\pm\left(20b_5\left(r-r_0\right)+9b_3^2\right)^{\frac{1}{2}}\right)+\mathcal{O}(\frac1h). 
		\label{eq:U_alpha_beta}
	\end{flalign}
	The boundaries of the pinning region are then given by	
	\begin{eqnarray}
	r_{\alpha,\beta}&\approx&\hat{r}_{\alpha,\beta}\nonumber\\
	&&-\frac{2}{h^2}\left(2+\frac{\sqrt{5b_3+5\sqrt{4b_5\left(\hat{r}_{\alpha,\beta}-1\right)+b_3^2}}}{\sqrt{3b_3+\sqrt{20b_5\left(\hat{r}_{\alpha,\beta}-1\right)+9b_3^2}}}\right),\nonumber\\
	\label{eq:r_alpha_beta}
	\end{eqnarray}	
	\begin{figure}[t!]
		\centering
		{\includegraphics[scale=0.56]{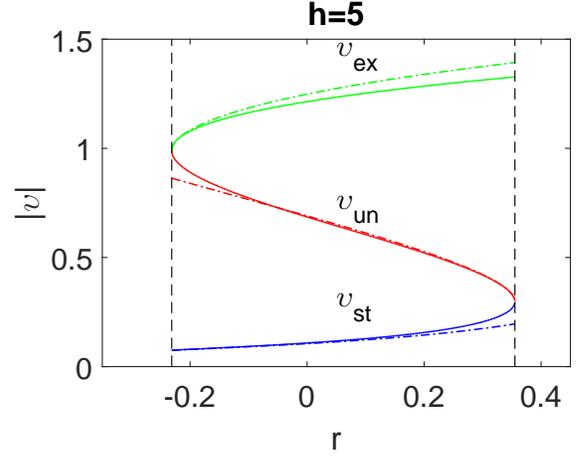}}
		\caption{Comparisons of the roots of \eqref{eq:poly_snake} that are related to snaking obtained numerically (solid lines) and the approximations \eqref{eq:U_st}, \eqref{eq:U_un}, and \eqref{eq:U_ex} (dashed lines). Here, $h=5$.
		}\label{fig:u_roots_approx}
	\end{figure}

	with
	\begin{eqnarray}
	\hat{r}_{\alpha}=1-\frac{b_3^2}{4b_5}, \quad
	\hat{r}_{\beta}=1\label{eq:pinning_bound}
	\end{eqnarray}

	Comparisons between the numerical results and the approximations above are shown in Figs.~\ref{fig:prb3_2}, \ref{fig:prb3_55}, and \ref{fig:prhg2}, where we can see that in general the approximation $r_{\alpha,\beta}$ gives good results particularly for the left boundary.

	{We can also asymptotically }obtain the three particular roots to be given by	
	\begin{eqnarray}
	\upsilon_{st} &\approx& \frac{\hat{\varepsilon}\left(2h^2-5\right)}{h^4\left(r-r_0\right)-4 h^2+10},\label{eq:U_st}\\
	\upsilon_{un}& =&\upsilon_\beta-\sqrt{\frac{\left(r-r_\beta\right)}{10\upsilon_\beta^2b_5-3b_3}}+\mathcal{O}\left(r-r_\beta\right),\label{eq:U_un}\\
	\upsilon_{ex} &=& \upsilon_\alpha+\sqrt{\frac{\left(r-r_\alpha\right)}{10\upsilon_\alpha^2b_5-3b_3}}+\mathcal{O}\left(r-r_\alpha\right).\label{eq:U_ex}
	\end{eqnarray}	
	
	Comparisons between the numerically computed roots of \eqref{eq:poly_snake} relevant to snaking and the approximations \eqref{eq:U_st}, \eqref{eq:U_un}, and \eqref{eq:U_ex} are shown in Fig.\ \ref{fig:u_roots_approx}.
	
\begin{figure*}[htbp!]
		\centering
		\subfigure[]{\includegraphics[scale=0.39]{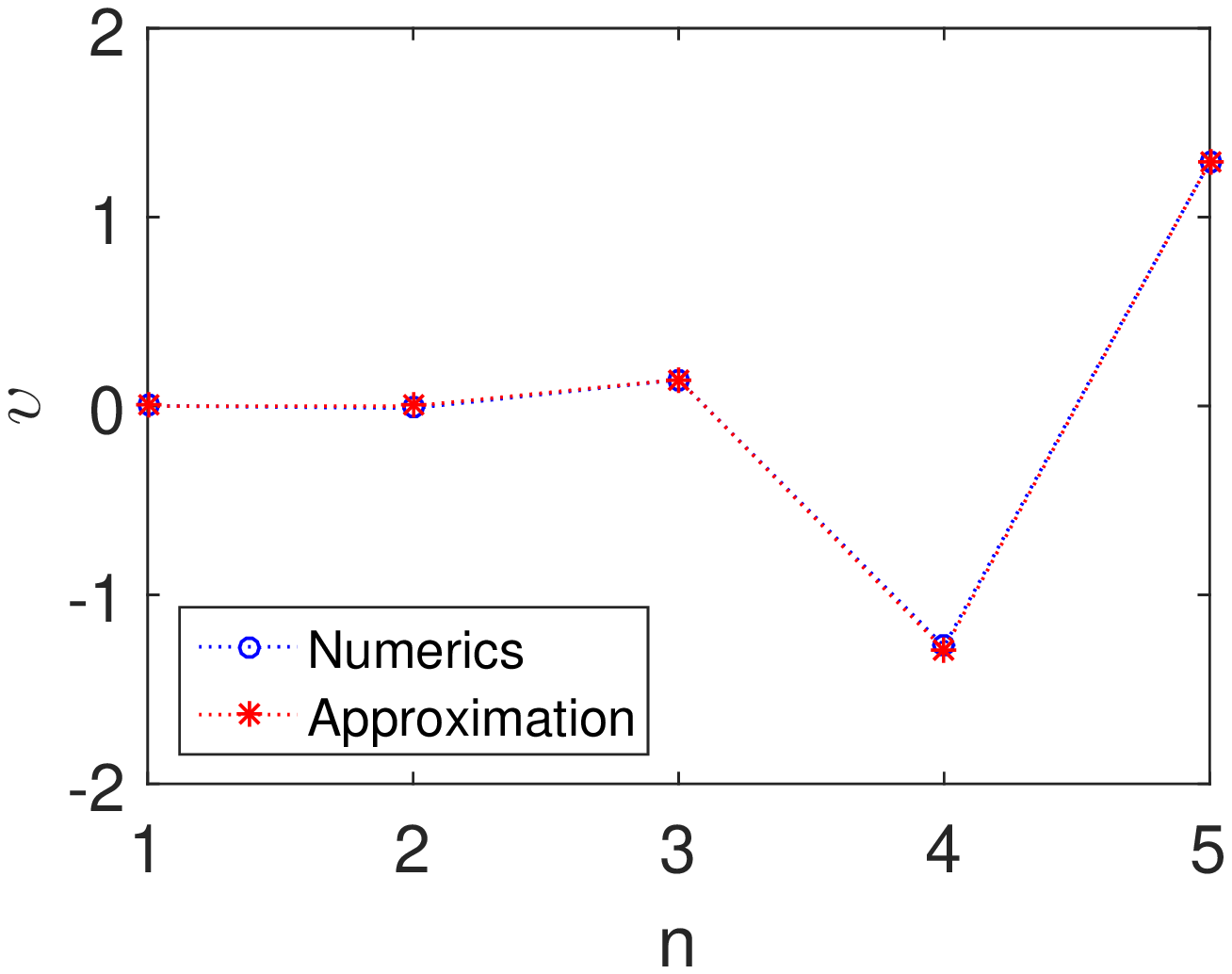}\label{subfig:prof_compare_num_approx_h5_st}}
		\subfigure[]{\includegraphics[scale=0.39]{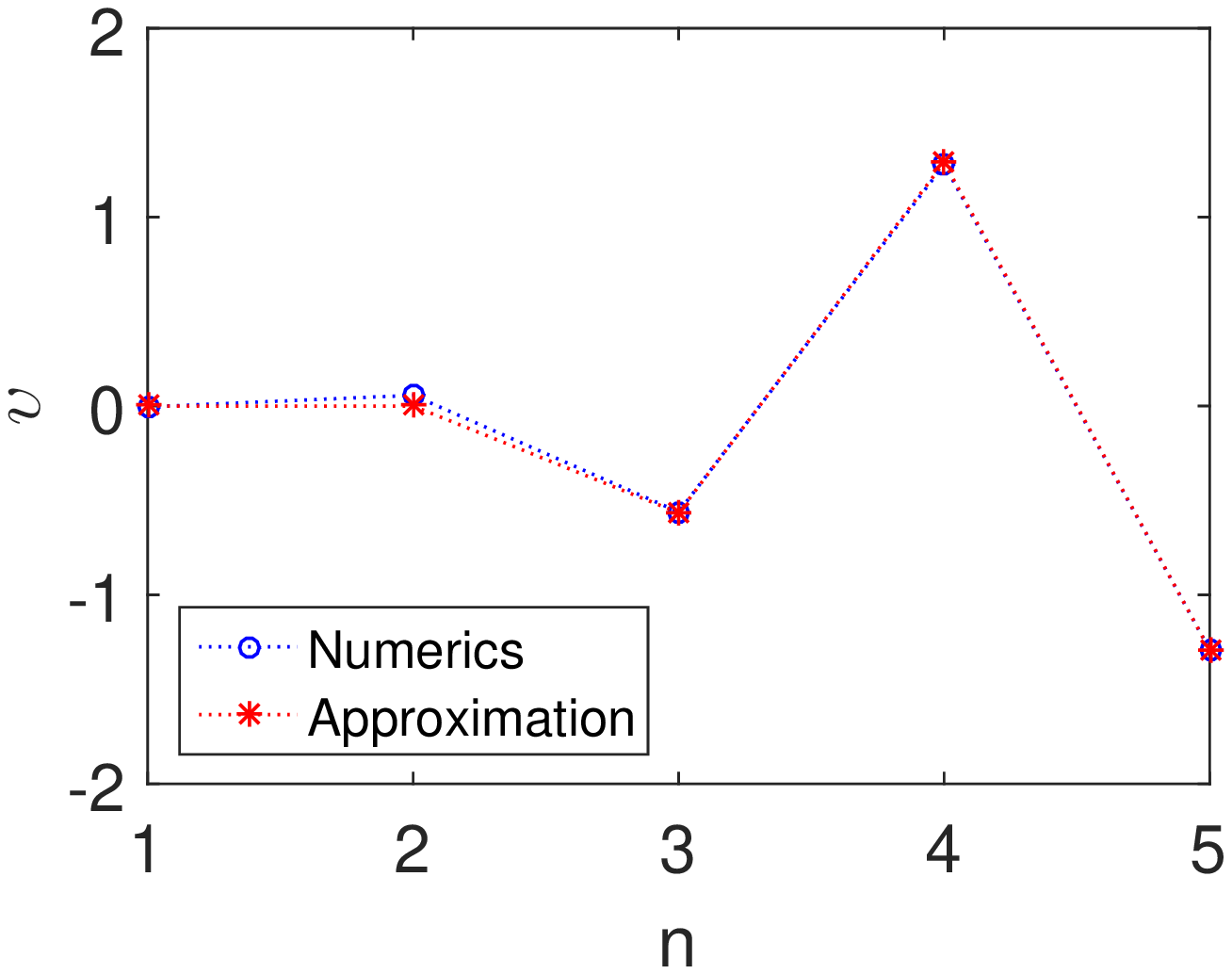}\label{subfig:prof_compare_num_approx_h5_un}}
		\subfigure[]{\includegraphics[scale=0.39]{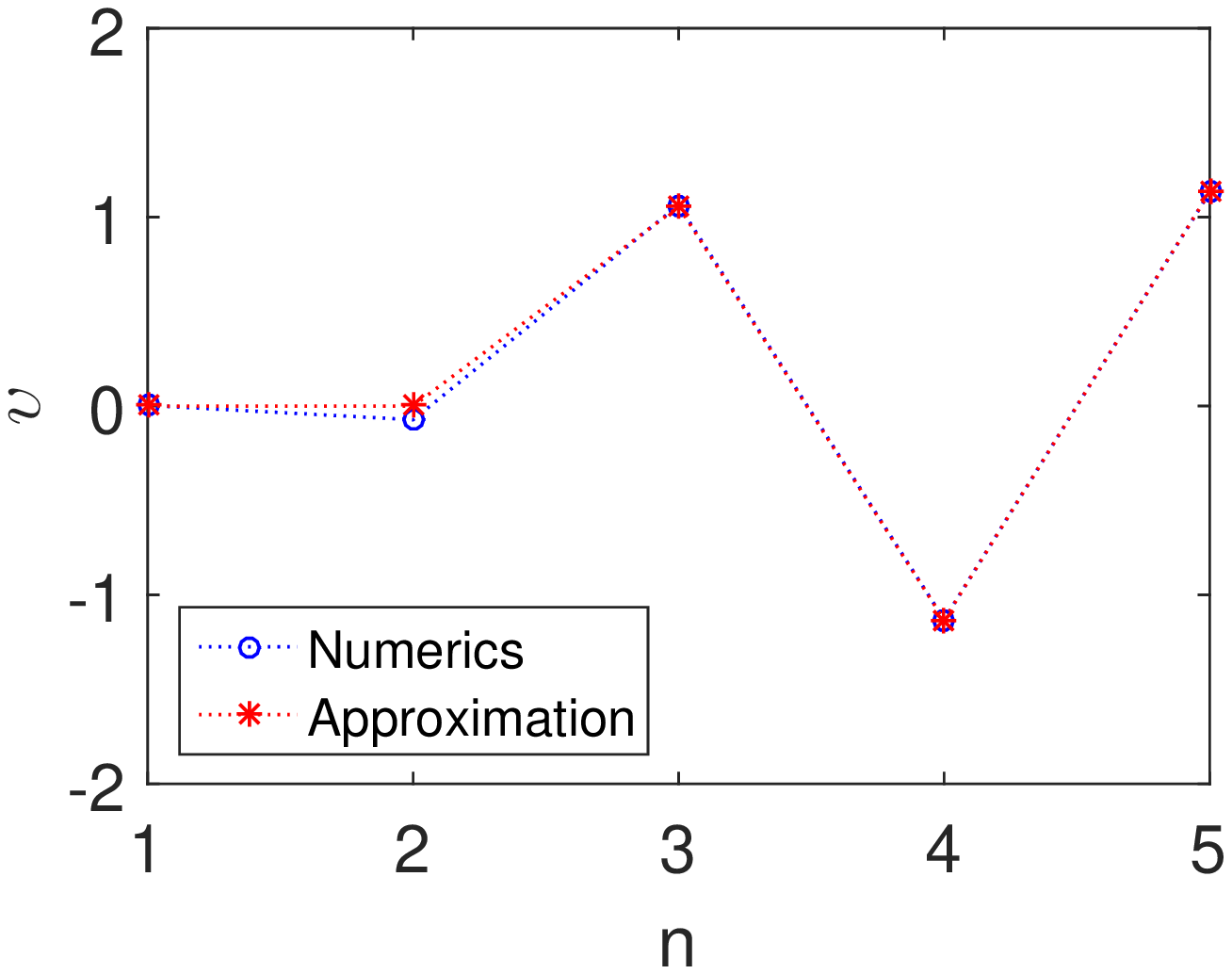}\label{subfig:prof_compare_num_approx_h5_ex}}
		\subfigure[]{\includegraphics[scale=0.39]{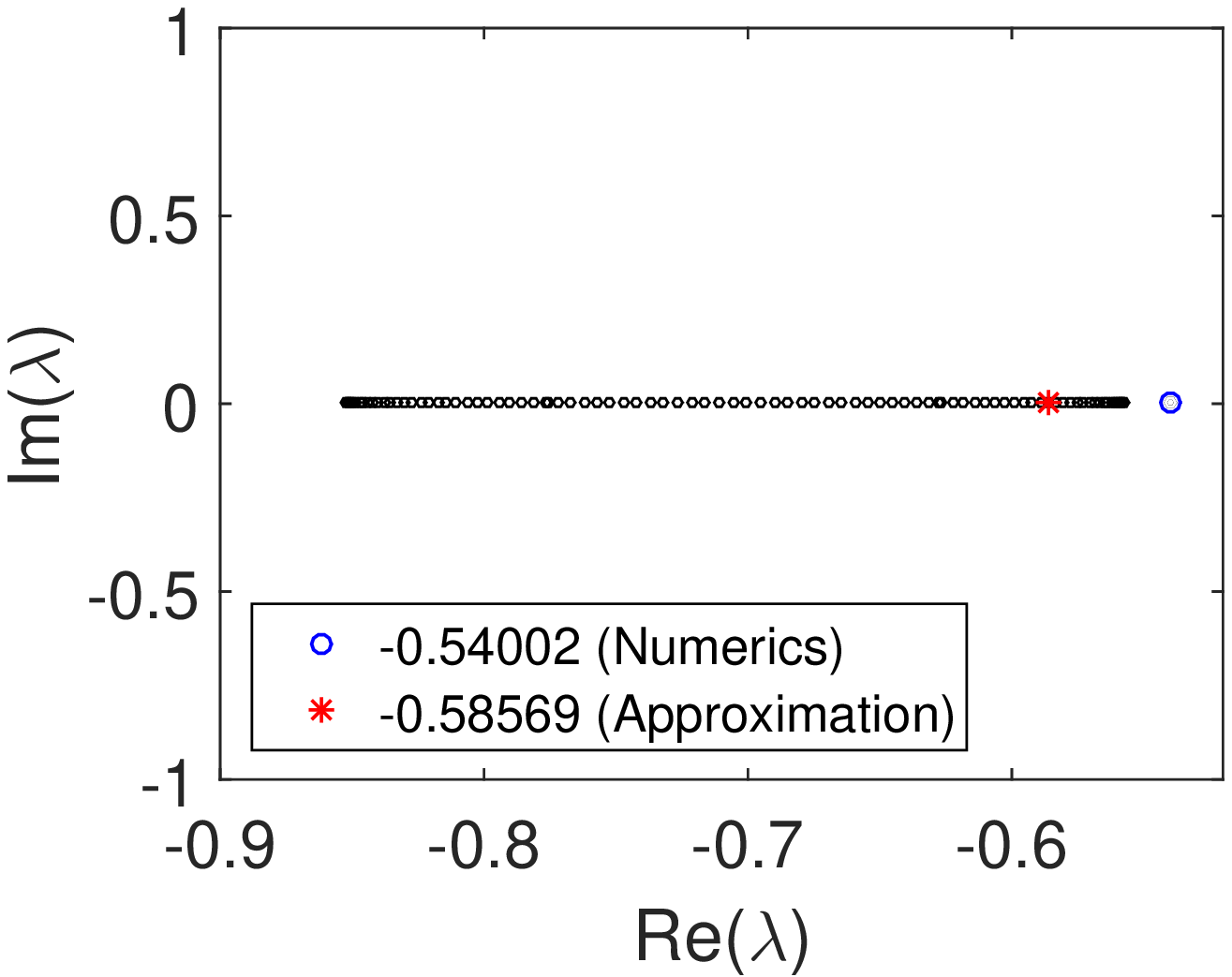}\label{subfig:eig_compare_num_approx_h5_st}}
		\subfigure[]{\includegraphics[scale=0.39]{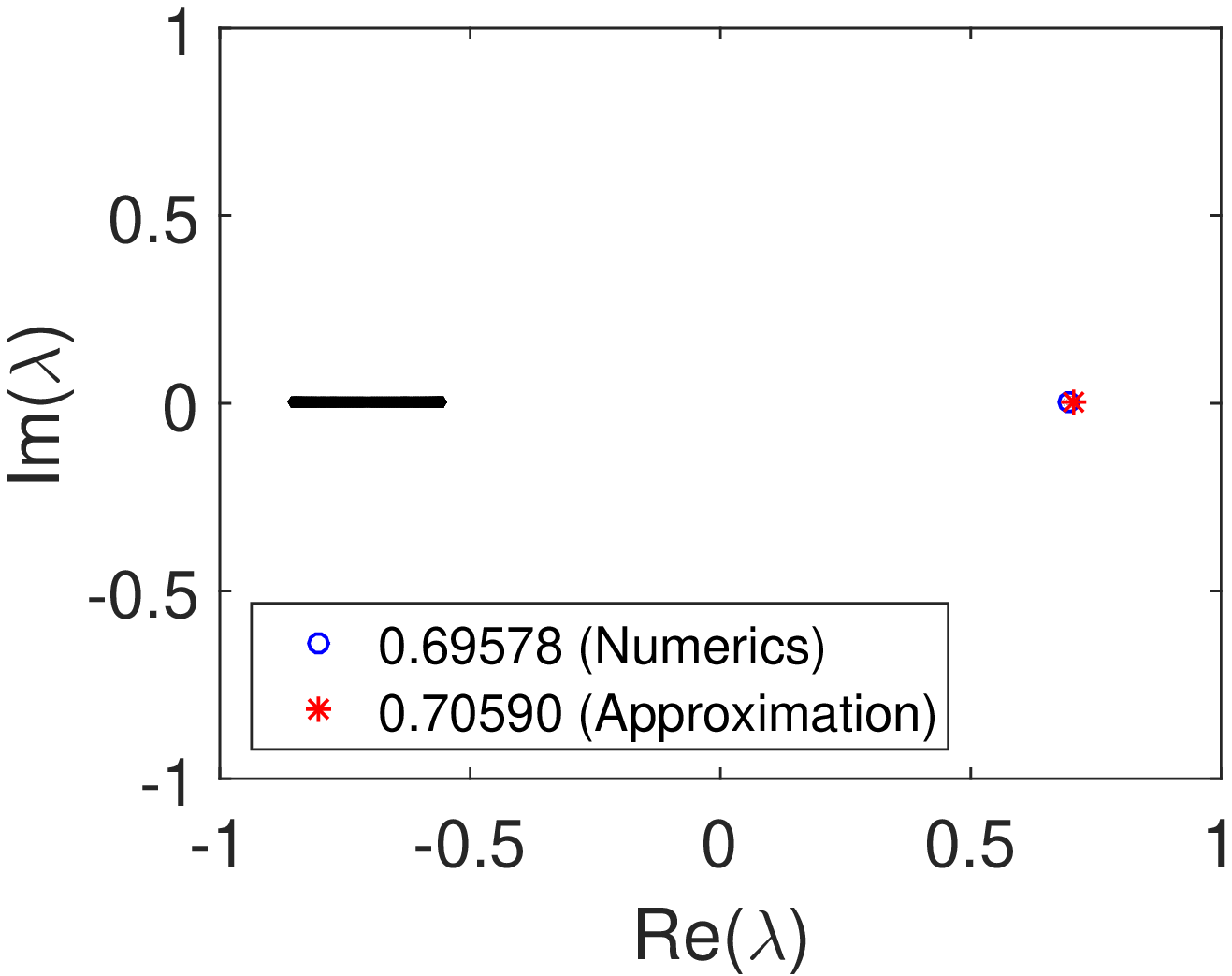}\label{subfig:eig_compare_num_approx_h5_un}}	
		\subfigure[]{\includegraphics[scale=0.39]{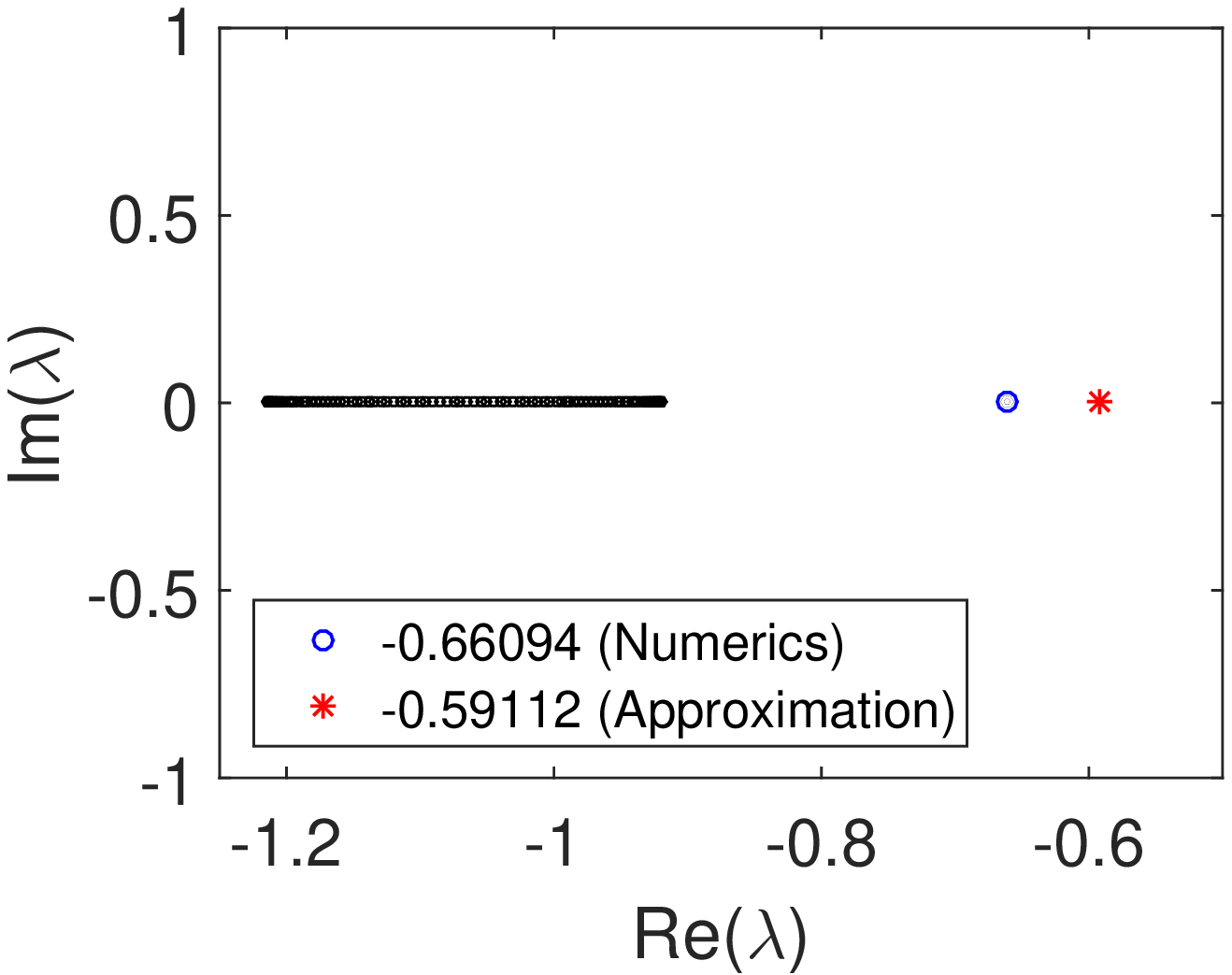}\label{subfig:eig_compare_num_approx_h5_ex}}	
		\caption{(a)-(c) Comparisons between the numerically obtained localized solutions of the discrete Swift-Hohenberg equation \eqref{eq:dSwift-Hohenberg} and the one-active site approximation \eqref{eq:coupling_sh_assume} for $h=5$. (d)-(f) The corresponding {spectrum} of the localized solutions in the top panels that are computed numerically and the eigenvalue \eqref{eigap} approximating the critical spectrum.
		}\label{fig:compare_prof_eig}
	\end{figure*}

	We compare in Fig.\ \ref{subfig:prof_compare_num_approx_h5_st}, \ref{subfig:prof_compare_num_approx_h5_un}, and \ref{subfig:prof_compare_num_approx_h5_ex} the numerical results obtained from the solution of the full system 
	and the approximations \eqref{eq:coupling_sh_assume} using roots of the one-active site polynomial \eqref{eq:poly_snake}. 
	One can see that the approximations are good.

	Next, we will show that the one-active site approximation can also be used to approximate the critical eigenvalue of localized solutions in the pinning region. This is obtained from realising that the dynamics of the active site will satisfy the equation $\upsilon_t=P_5(\upsilon)$. It is then immediate that the eigenvalue will be given by the linear eigenvalue problem
	\begin{eqnarray}
	\lambda\upsilon=\left.\frac{d}{d\upsilon}P_5(\upsilon)\right|_{\upsilon=\upsilon_{st,un,ex}}\upsilon,
	\end{eqnarray}
	i.e.,\
	\begin{eqnarray}
		\lambda_{st,un,ex}(r)&=&-5b_5\upsilon_{st,un,ex}^4+3b_3\upsilon_{st,un,ex}^2\nonumber\\
		&&+\left(r-1+\dfrac{4}{h^2}-\dfrac{6}{h^4}\right).
		\label{eigap}
	\end{eqnarray}
	Figures\ \ref{subfig:eig_compare_num_approx_h5_st}, \ref{subfig:eig_compare_num_approx_h5_un}, and \ref{subfig:eig_compare_num_approx_h5_ex} show numerically computed spectrum of the profiles in Figs.\ \ref{subfig:prof_compare_num_approx_h5_st}, \ref{subfig:prof_compare_num_approx_h5_un}, and \ref{subfig:prof_compare_num_approx_h5_ex} and our approximation \eqref{eigap}, where rather excellent agreement is obtained.
	
	\section{Conclusion}
	\label{sec6}

	We have considered a discrete Swift-Hohenberg equation that is obtained from discretizing the spatial derivatives of the continuous one. We have studied time-independent solutions, namely, uniform, periodic, and localized solutions and their (in)stabilities, from which we concluded that in terms of the discretization parameter $h$, the equation can be distinguished into three different regions, i.e.,\ $0<h<1$, $1\leq h<2$, and $h\geq2$. In the first interval, the uniform, the periodic and the localized solutions of the discrete Swift-Hohenberg equation have similar properties with the continuous case. As a direct consequence, our study indicates that to solve the (continuous) Swift-Hohenberg equation numerically using finite central differences, it can be sufficient to use relatively large $h<1$.	
	
	As the discretization parameter becomes larger, features different from the continuous counterpart may emerge, such as instability of localized solutions for both phase $\phi=0$ and $\pi/2$ for the same parameter values, extra bifurcation curves for the periodic and localized solutions, and multiple Maxwell points. Moreover, one may also obtain a snaking structure in the bifurcation diagram of periodic solutions, that does not exist in the continuous limit, as well as complicated snaking structures for localized solutions.

	Analytical approximations have been developed for the periodic and the localized solutions. The periodic solution amplitudes have been determined using variational methods, while the localized solutions have been approximated using asymptotic analysis.

	The boundaries of the pinning region, i.e.,\ of the homoclinic snaking that is associated with the localized solutions, have been studied numerically as well as analytically by developing a one-active site approximation. We have shown that the approximation can also be used to approximate the critical eigenvalue of a localized solution. Comparisons of the analytical results and the numerics show good agreement.

	\begin{figure}[htbp!]
	\centering
	\includegraphics[scale=0.55]{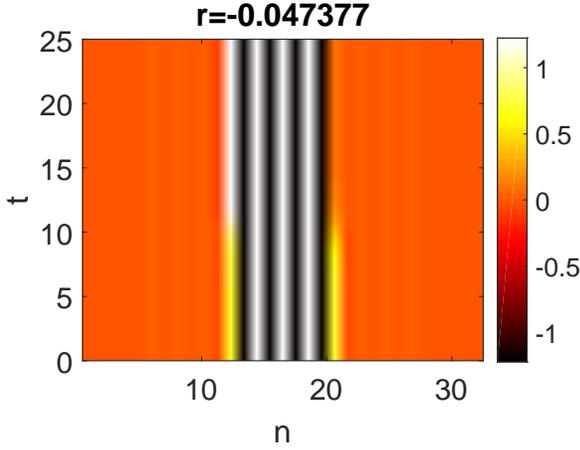}
	\caption{Time dynamics of the unstable solution shown in Fig.\ \ref{subfig:prof_h5_b}, that corresponds to point b in Fig.\ \ref{subfig:Ls_h5}. The symmetric solution evolves into an antisymmetric one, which corresponds to a point on the stable branch right above point b. 
	}
	\label{fig:prof_h5_b_perturb}
\end{figure}

	In this work, we mainly only considered time-independent solutions, where we determined their local (in)stability from computing the spectrum of their corresponding linear differential operator. The typical time evolution of the unstable solutions, which is rather related to global dynamics, is depicted in Fig.\ \ref{fig:prof_h5_b_perturb}, where an unstable solution would settle into a stable neighboring one.%
	
	Discussing time-dependent solutions, 
		it is interesting to study the effect of parametric time-periodic forcing \mbox{\cite{Gandhi2015}} to the snaking behavior in discrete systems, which is addressed for future work. The mechanism for snaking or non-snaking in discrete systems (in the continuous case, it is discussed in \mbox{\cite{Avitabile2010}}) is also proposed to be studied in the future.

	\section*{Acknowledgement}
	
	RK gratefully acknowledges financial support from Lembaga Pengelolaan Dana Pendidikan (Indonesia Endowment Fund for Education) (Grant No.-Ref:S-34/LPDP.3/2017). The authors gratefully acknowledge the two anonymous reviewers for their careful reading. 
	
\appendix

	\section{Asymptotic expansion of localized solutions}
\label{sec:asymp}
Defining new variables \cite{Yi2009}
\begin{equation*}
	X=\epsilon n,  \quad T=\epsilon^2t,\quad r=\epsilon^2 r_1+r_0,
	\label{eq:asymp_var}
\end{equation*}
and writing
\begin{equation}
\begin{array}{ccl}
u_n(t)&=&\epsilon e^{i \psi}F\left(X,\tau,T\right)+\epsilon^2 G_0 \left(X,\tau,T\right)\\
&&+\epsilon^2 e^{i \psi}G_1 \left(X,\tau,T\right)+\epsilon^2 e^{2i \psi}G_2 \left(X,\tau,T\right)\\
&&+\epsilon^3 H_0 \left(X,\tau,T\right)+\epsilon^3 e^{i \psi}H_1 \left(X,\tau,T\right)\\
&&+\ldots+\text{c.c.},
\end{array}
\label{eq:asypm_exp_c}
\end{equation}
{where  $\psi=khn$ with $k$ being the wavenumber of the carrier wave \eqref{eq:khl2}, \eqref{eq:khg2}, and c.c.\ denotes the complex conjugation, we obtain}
\begin{eqnarray}
u_{n\pm j}(t)&=&\epsilon e^{i \psi_{n\pm j}}\left[F\pm j \epsilon\frac{\partial F}{\partial X}+\left(j\epsilon\right)^2\frac{1}{2}\frac{\partial^2 F}{\partial X^2}\pm\ldots\right]\nonumber\\
&&+\epsilon^2 \left[G_0\pm j\epsilon\frac{\partial G_0}{\partial X}+\left(j\epsilon\right)^2\frac{1}{2}\frac{\partial^2 G_0}{\partial X^2}\pm\ldots\right]\nonumber\\
&&+\epsilon^2 e^{i \psi_{n\pm j}}\left[G_1\pm j\epsilon\frac{\partial G_1}{\partial X}+\left(j\epsilon\right)^2\frac{1}{2}\frac{\partial^2 G_1}{\partial X^2}\pm\ldots\right]\nonumber\\
&&+\ldots+\text{c.c.}
\label{eq:taylor_pm1}
\end{eqnarray}
and
\begin{eqnarray}
\frac{d}{d t}=\frac{\partial}{\partial t}+\epsilon^2\frac{\partial}{\partial T}.
\label{eq:total_diff}
\end{eqnarray}

{Next, we substitute Eqs.\ \mbox{\eqref{eq:asypm_exp_c}} and \mbox{\eqref{eq:taylor_pm1}} into the discrete Swift-Hohenberg equation \mbox{\eqref{eq:dSwift-Hohenberg}} and equate the coefficients of each harmonic in $\psi$ at each order of $\epsilon$.} 

At $\mathcal{O}\left(\epsilon e^{{i\psi}}\right)$, we have
\begin{eqnarray}
\begin{aligned}
&\left[1-r_{{0}}+4\left(\frac {\cos \left( kh \right) -1}{{h}^{2}}\right)\right.
\left.\left( 1+{\frac {\cos \left( kh \right) -1}{{h}^{2}}} \right) \right]F=0.
\end{aligned}\nonumber\\
\label{eq:eps1_psi1}
\end{eqnarray}
{Because $F$ cannot be zero, its coefficient must vanish, which is satisfied for $k$ and $r_0$ given by Eqs.\ \eqref{eq:khl2} and \eqref{eq:r0_hl2}, or \eqref{eq:khg2} and \eqref{eq:r0_hg2}, respectively.

At $\mathcal{O}\left(\epsilon^2 e^{{i\psi}}\right)$, we obtain
\begin{eqnarray}
\left[{\frac {4\,i\sin \left( kh \right) }{{h}^{4}}} \left( 2\left(\cos \left( kh \right) - 1\right)+{h}^{2}\right) \right]=0.
\label{eq:vel_wavnum}
\end{eqnarray}

At $\mathcal{O}\left(\epsilon^3 e^{{i\psi}}\right)$, we obtain
\begin{eqnarray}
{} F_T  &=&{A}G_{1X}+{C}F_{XX}+3\,b_{{3}}\overline{F}{F}^{2}+r_{{1}}F\nonumber\\
&&-\left[1-r_{{0}}+4\left(\frac {\cos \left( kh \right) -1}{{h}^{2}}\right)\right.\nonumber\\
&&\left.\left( 1+{\frac {\cos \left( kh \right) -1}{{h}^{2}}} \right) \right]H_1,\nonumber\\
\label{eq:eps3_psi1}
\end{eqnarray}
where 
\begin{eqnarray*}
	A &=&  -\left[{\frac {4\,i\sin \left( kh \right) }{{h}^{4}}} \left( 2\left(\cos \left( kh \right) - 1\right)+{h}^{2}\right) \right]\\
	C &=& -2\left[\frac{ \cos \left( kh \right)  \left( {h}^{2}-2 \right)+2\cos \left( 2kh \right)}{h^4}\right].
\end{eqnarray*}
{By using \mbox{\eqref{eq:eps1_psi1}} and \mbox{\eqref{eq:vel_wavnum}}, we can eliminate the coefficient of $H_1$ and obtain Ginzburg-Landau equation for $F$
}
\begin{eqnarray}
{F_T =CF_{XX}+3b_{{3}}|{F}|^2{F}+r_{{1}}F.}
\label{eq:eps3_psi1_el1}
\end{eqnarray}
Because we focus on the time-independent system $F_T=0$, we have
\begin{eqnarray}
C\,F_{XX}+3\,b_{{3}}|{F}|^2{F}+r_{{1}}F=0,
\label{eq:F_ti}
\end{eqnarray}
where
\begin{eqnarray}
C&=&{-\frac { \left( {h}^{2}-4 \right) }{{h}^{2}}}\label{eq:C_hl2}
\end{eqnarray}
and
\begin{eqnarray}
C&=&\,{\frac {2\left({h}^{2}-4\right)}{{h}^{4}}}\label{eq:C_hg2}
\end{eqnarray}
for $h\leq2$ and $h>2$, respectively.

The uniform solution of equation \eqref{eq:F_ti} is
\begin{eqnarray}
F(X)=\left(-\frac{r_1}{3\,b_3}\right)^{\frac{1}{2}}e^{i \phi}, 
\label{eq:F_zero}
\end{eqnarray}
corresponding to spatially periodic states with period $P$ near $r=0$
\begin{eqnarray}
{u}_{P,n}=2\left(\frac{(r_0-r)}{3\,b_3}\right)^{\frac{1}{2}}\cos\left(khn+\phi\right)+\mathcal{O}(r-r_0).\nonumber\\
\label{eq:Un_zero}
\end{eqnarray}

Localized states satisfying $F\rightarrow0$ as $X\rightarrow\pm\infty$ are given by
\begin{eqnarray}
F(X)=\left(-\frac{2r_1}{3b_3}\right)^{\frac{1}{2}}\sech\left(X\left(\frac{r_1}{C} \right)^{\frac{1}{2}}\right)e^{i \phi}, 
\label{eq:F_loc}
\end{eqnarray}
that using \eqref{eq:asypm_exp_c} lead to the solution \eqref{eq:Un_loc}.



\begin{thebibliography}{999}
		\bibitem{Woods2006} P. D. Woods and A. R. Champneys, \href{http://dx.doi.org/10.1016/S0167-2789(98)00309-1}{Physica D \textbf{129}, 147 (1999)}.
		\bibitem{Knobloch2008} E. Knobloch, \href{http://dx.doi.org/10.1088/0951-7715/21/4/T02}{Nonlinearity \textbf{21}, T45 (2008)}.
				\bibitem{Dawes2010} J. H. P. Dawes, \href{http://dx.doi.org/10.1098/rsta.2010.0057}{Philos. Trans. R. Soc. Lond. A \textbf{368}, 3519 (2010)}.
		\bibitem{Purwins2010} H.-G. Purwins, H. U. Bo\"edeker, and Sh. Amiranashvili, \href{http://dx.doi.org/10.1080/00018732.2010.498228}{Adv. Phys. \textbf{59}, 485 (2010)}.
		\bibitem{Swift1977} J. Swift and P. C. Hohenberg, \href{https://doi.org/10.1103/PhysRevA.15.319}{Phys. Rev. A \textbf{15}, 319 (1977)}.
\bibitem{Getling1998} A. V. Getling, \textit{Rayleigh-Bénard Convection}, Advanced Series in Nonlinear Dynamics, Vol. 11 (World Scientific, Singapore, 1998).
		\bibitem{Budd2001} C. J. Budd, G. W. Hunt, and R. Kuske, \href{http://dx.doi.org/10.1098/rspa.2001.0843}{Proc. R. Soc. London A \textbf{457}, 2935 (2001)}.
		\bibitem{Hunt2000} G. W. Hunt, M. A. Peletier, A. R. Champneys, P. D. Woods, M. A. Wadee, C. J. Budd, and G. J. Lord, \href{http://dx.doi.org/\%2010.1023/A:1008398006403}{Nonlinear Dyn. \textbf{21}, 3 (2000)}.
		\bibitem{Beck2009} M. Beck, J. Knobloch, D. J. B. Lloyd, B. Sandstede, and T. Wagenknecht, \href{http://dx.doi.org/10.1137/080713306}{SIAM J. Math. Anal. \textbf{41}(3), 936 (2009)}.
		\bibitem{Budd2005} C. J. Budd and R. Kuske, \href{http://dx.doi.org/10.1016/j.physd.2005.06.009}{Physica D \textbf{208}, 73 (2005)}.
		\bibitem{Kozyreff2006} G. Kozyreff and S. J. Chapman, \href{https://doi.org/10.1103/PhysRevLett.97.044502}{Phys. Rev. Lett. \textbf{97}, 044502 (2006)}.
		\bibitem{Burke2006} J. Burke, E. Knobloch, \href{https://doi.org/10.1103/PhysRevE.73.056211}{Phys. Rev. E \textbf{73}, 056211 (2006)}.
		\bibitem{Burke2007} J. Burke and E. Knobloch, \href{http://dx.doi.org/10.1063/1.2746816}{Chaos \textbf{17}, 037102 (2007)}.
		\bibitem{Burke2007a} J. Burke and E. Knobloch, \href{Phys. Lett. A 360, 681 (2007)}{Phys. Lett. A \textbf{360}, 681 (2007)}.
		\bibitem{Clerc2005} M. G. Clerc and C. Falcon, \href{http://dx.doi.org/10.1016/j.physa.2005.05.011}{Physica A \textbf{356}, 48 (2005)}.
\bibitem{Barbay2008} S. Barbay, X. Hachair, T. Elsass, I. Sagnes, and R. Kuszelewicz, \href{https://doi.org/10.1103/PhysRevLett.101.253902}{Phys. Rev. Lett. \textbf{101}, 253902 (2008)}.
		\bibitem{Haudin2011} F. Haudin, R. G. Rojas, U. Bortolozzo, S. Residori, and M. G. Clerc, \href{https://doi.org/10.1103/PhysRevLett.107.264101}{Phys. Rev. Lett. \textbf{107}, 264101 (2011)}.
		\bibitem{Lloyd2015} D. J. B. Lloyd, C. Gollwitzer, I. Rehberg, and R. Richter,	\href{https://doi.org/10.1017/jfm.2015.565}{J. Fluid Mech. \textbf{783}, 283 (2015)}.
		\bibitem{Pomeau1986} Y. Pomeau, \href{http://dx.doi.org/10.1016/0167-2789(86)90104-1}{Physica D \textbf{23}, 3 (1986)}.
		\bibitem{Bensimon1988} D. Bensimon, B. I. Shraiman, and V. Croquette, \href{https://doi.org/10.1103/PhysRevA.38.5461}{Phys. Rev. A \textbf{38}, 5461 (1988)}.
		\bibitem{Sakaguchi1996} H. Sakaguchi and H. R. Brand, \href{http://dx.doi.org/10.1016/0167-2789(96)00077-2}{Physica D \textbf{97}, 274 (1996)}.
		\bibitem{Chapman2009} S. J. Chapman and G. Kozyreff, \href{http://dx.doi.org/10.1016/j.physd.2008.10.005}{Physica D \textbf{238}, 319 (2009)}.
		\bibitem{Dean2011} A. D. Dean, P. C. Matthews, S. M. Cox, and J. R. King. \href{http://dx.doi.org/10.1088/0951-7715/24/12/003}{Nonlinearity \textbf{24}, 3323 (2011)}.
		\bibitem{Susanto2011} H. Susanto and P. C. Matthews, \href{https://doi.org/10.1103/PhysRevE.83.035201}{Phys. Rev. E \textbf{83}, 035201(R) (2011)}.
		\bibitem{Matthews2011} P. C. Matthews and H. Susanto, \href{https://doi.org/10.1103/PhysRevE.84.066207}{Phys. Rev. E \textbf{84}, 066207 (2011)}.
		\bibitem{Carretero-Gonzalez2006} R. Carretero-Gonzalez, J. D. Talley, C. Chong, and B. A. Malomed, \href{http://dx.doi.org/10.1016/j.physd.2006.01.022}{Physica D \textbf{216}, 77 (2006)}.
		\bibitem{Chong2009} C. Chong, R. Carretero-Gonzalez, B. A. Malomed and P. G. Kevrekidis, \href{http://dx.doi.org/10.1016/j.physd.2008.10.002}{Physica D \textbf{238}, 126 (2009)}.
		\bibitem{Chong2011} C. Chong and D. E. Pelinovsky, \href{http://dx.doi.org/10.3934/dcdss.2011.4.1019}{Disc. Cont. Dyn. Sys. S \textbf{4}, 1019 (2011)}.
		\bibitem{Taylor2010} C. Taylor and J. H. P. Dawes, \href{http://dx.doi.org/10.1016/j.physleta.2010.10.010}{Phys. Lett. A \textbf{375}, 14 (2010)}.
		\bibitem{Yulin2008} A. V. Yulin, A. R. Champneys, and D. V. Skryabin, \href{https://doi.org/10.1103/PhysRevA.78.011804}{Phys. Rev. A \textbf{78}, 011804(R) (2008)}.
		\bibitem{Yulin2010} A. V. Yulin and A. R. Champneys, \href{http://dx.doi.org/10.1137/080734297}{SIAM J. Appl. Dyn. Syst. \textbf{9}(2), 391 (2010)}.
		\bibitem{Clerc2011} M. G. Clerc, R .G. Elias and R. G. Rojas, \href{https://doi.org/10.1098/rsta.2010.0255}{Phil. Trans. Roy. Soc. A \textbf{369}, 412 (2011)}.
		\bibitem{mccu16} N.\ McCullen and T.\ Wagenknecht, Scientific Reports 6, 27397 (2016).
		\bibitem{Dean2015} A. D. Dean, P. C. Matthews, S. M. Cox, and J. R. King, \href{http://dx.doi.org/10.1137/140966897}{SIAM J. Appl. Dyn. Syst. \textbf{14}(1), 481 (2015)}.
		\bibitem{Peletier2004} L. A. Peletier and J. A. Rodr\'iguez (unpublished).
		\bibitem{Rodriguez2004} J. A. Rodr\'guez, {\textit{Patterns described by Discrete and Continuous Dynamical systems}}, Universiteit Leiden, Ph.D. thesis (2004).
		\bibitem{Collet1998} P. Collet, \href{https://doi.org/10.1023/A:1023212925677}{J. Stat. Phys \textbf{90}, 1075 (1998)}.		
		\bibitem{hunt11} J.K. Hunter and B. Nachtergaele, \textit{Applied Analysis} (World Scientific, River Edge, NJ, 2001).
		\bibitem{Oppenheim1983}  {A. V. Oppenheim, A. S. Willsky, and I. T. Young, \mbox{\textit{Signals and Systems}}, Prentice Hall (1983).}
		\bibitem{Seydel1994} R. Seydel, \textit{Practical Bifurcation and Stability Analysis: From Equilibrium to Chaos}, 2nd ed., Interdisciplinary Applied Mathematics	Vol. 5 (Springer-Verlag, Berlin, 1994).
		\bibitem{add1} G. Kozyreff, P. Assemat, and S. J. Chapman, \href{https://journals.aps.org/prl/abstract/10.1103/PhysRevLett.103.164501}{Phys. Rev. Lett. 103, 164501 (2009).} 
		\bibitem{Gandhi2015} P. Gandhi, C. Beaume, and E. Knobloch, \href{http://dx.doi.org/10.1137/14099468X}{SIAM J. Appl. Dyn. Syst. \textbf{14}(2), 860 (2015)}.
		\bibitem{Avitabile2010} D. Avitabile, D. Lloyd, J. Burke, E. Knobloch, and	B. Sandstede, \href{http://dx.doi.org/10.1137/100782747}{SIAM J. Appl. Dyn. Syst. \textbf{9}(3), 704 (2010)}.
		\bibitem{Yi2009} X. Yi, J. A. D. Wattis, H. Susanto, and L. J. Cummings, \href{http://dx.doi.org/10.1088/1751-8113/42/35/355207}{J. Phys. A \textbf{42}, 355207 (2009)}.
	\end{thebibliography}
\end{document}